# Stable Adaptive Control Using New Critic Designs

Paul J. Werbos
Room 675, National Science Foundation[*] *
Arlington, Virginia, USA

## Abstract

Classical adaptive control proves total-system stability for control of linear plants, but only for plants meeting very restrictive assumptions. Approximate Dynamic Programming (ADP) has the potential, in principle, to ensure stability without such tight restrictions. It also offers nonlinear and neural extensions for optimal control, with empirically supported links to what is seen in the brain. However, the relevant ADP methods in use today -- TD, HDP, DHP, GDHP -- and the Galerkin-based versions of these all have serious limitations when used here as parallel distributed real-time learning systems; either they do not possess quadratic unconditional stability (to be defined) or they lead to incorrect results in the stochastic case. (ADAC or Q-learning designs do not help.) After explaining these conclusions, this paper describes new ADP designs which overcome these limitations. It also addresses the Generalized Moving Target problem, a common family of static optimization problems, and describes a way to stabilize large-scale economic equilibrium models, such as the old long-term energy model of DOE.

## 1. Introduction

The main goal of this paper is to present the new research and new designs which emerged from an effort to find new connections between certain streams of classical control theory[1-6] and biologically-inspired intelligent control [7-15]. As one might expect, the main benefits of this work are to these two areas. However, as a byproduct, new designs were also found for the Generalized Moving Target (GMT) problem [16,17], which is defined as follows: for a function $E(\underline{v}, \underline{w})$, find two vectors $\underline{v}$ and $\underline{w}$ such that the following two conditions are met:
(1) E is minimized with respect to $\underline{v}$ for $\underline{w}$ fixed; (2) $\underline{w} = \underline{v}$. Likewise, a new insight -- simple, but important in practical terms -- was found, regarding the task of reliable convergence for large-scale energy-economic models (or other models) made up of large numbers of algebraic equations which must be solved simultaneously [16]. This might even have implications for the task of stabilizing actual systems of economic exchange.

      The work reported here is intended to be the first step of a larger effort to improve the stability and performance of several classes of control design. These preliminary results address only two classes of control design: (1) "adaptive control," in the tradition of Narendra, which includes both linear (multiple-input multiple-output, MIMO) designs [1]
and nonlinear or neural extensions [18,19]; (2) learning-based Approximate Dynamic Programming (ADP) [7-15], which has sometimes been presented as a form of "reinforcement learning" [20-22], sometimes called "adaptive critics" [23] and sometimes called "neuro-dynamic programming"[24]. However, the results here also have implications for other streams of control theory such as robust nonlinear control [5,6], model-predictive control [2,8], and the development of automated tools to find Liapunov functions to stabilize nonlinear plants. For example, they may be relevant to the effort to develop piecewise-quadratic Liapunov functions for plants which can be approximated in piecewise-linear form.

---

[*] The views and the new designs described here are those of the author developed on personal time; they do not reflect the views of NSF. Many thanks are due to Dr. Ludmila Dolmatova and to Dr. Sam Leven, without whose help this would have been impossible.

Few readers will already be aware of all the connections and recent results in these fields. Therefore, this paper will begin with an extended discussion of the background to this new work. Section 2 will summarize the general motivation and context for the new work. It will discuss the current and potential implications of adaptive control and ADP, in conjunction with other control approaches, both in near-term engineering applications and in the longer-term effort to build truly intelligent systems for decision and control. Section 3
will review the standard forms of adaptive control discussed by Narendra, and the difficulty of ensuring stability even in the linear-quadratic case. It will show that there is a very close connection between these designs and a particular ADP design, HDP+BAC[8,22,25], which I actually developed back in 1971, as part of my PhD proposal[26]. In fact, when HDP+BAC is applied to the linear-quadratic case, it reduces down to a standard indirect adaptive control (IAC) design, <u>except that</u> there is a provision for adapting one extra matrix which, in principle, should permit much stronger stability guarantees. Roughly speaking, ordinary IAC designs are based on the minimization of tracking error, which is sometimes just:

$$\sum_i (x_i^{(ref)}(t+1) - x_i(t+1))^2 = \underline{e}(t+1)^T \underline{e}(t+1), \quad (1)$$

where the vector $\underline{x}^{(ref)}(t)$ represents the desired state of the plant or external environment at time t, where $\underline{x}(t)$ represents the actual state, and $\underline{e}(t)$ represents the gap or error between the two. In the same situation, HDP+BAC reduces to the same overall design, **except that** we minimize:

$$\hat{J}(t+1) = e(t+1)^T C e(t+1), \quad (2)$$

where C is a matrix of "Critic weights" to be adapted by HDP. In the nonlinear case, equation 2 is replaced by an artificial neural network (ANN), in order to allow $\hat{J}$ to
approximate any nonlinear function.

Section 4 will review ADP designs in general, and discuss their potential relevance. It will also include some new material on the continuous-time version of HDP and new ADP methods based on the use of the Galerkin approach [27]. Section 5 will then describe a longer-term strategy for developing less restrictive stable adaptive control and automated control design for linear and nonlinear systems; again, the work in this paper is intended only as a first step, to make it possible for other researchers to fulfill this larger strategy.
Section 5 will also try to anchor this proposed strategy to a larger strategy for achieving true intelligent control.

In theory, if the critic weights C or the Critic network converged to the right values (the values which satisfy the Hamilton-Jacobi-Bellman equation [4,5,24,28]), then the function $\hat{J}$ would serve as a Liapunov function guaranteed to stabilize the overall system, if the system is controllable. However, none of the established methods for adapting a Critic possess quadratic unconditional stability, even in the linear deterministic case. The variations of these methods based on the Galerkin approach to solving differential equations do possess unconditional stability in that case, but they converge to the wrong weights almost always in the linear stochastic case.

Sections 6 and 7 will substantiate these conclusions. Section 8 will discuss their general significance, links to similar results from other authors, and links to the Generalized Moving Target (GMT) problem and to models made up of systems of algebraic equations. Section 9 will propose several new methods which meet the tests of sections 6 and 7, and also discuss a new method in the previous literature (the "two sample method"), which also does so.

The notation in this paper will try to be consistent with the sources it draws upon, as much as possible. However, because of the difficulties in electronic document transmission, it will make two biased selections. I will simply write "C", "**b**" and W for the <u>estimated</u> values of certain sets of parameters or weights; I will write "C*" or "**b***" for their true values. But for the functions J and $\underline{\lambda}$, I will follow the usual three-fold convention of using the "J" and "$\underline{\lambda}$" for the true values, "J*" and "$\underline{\lambda}$*" for target values used in adaptation, and "$\hat{J}$" and "$\hat{\underline{\lambda}}$" for the estimated values. I will use "$\partial_t \underline{x}$" for the state vector in linear adaptive control, rather than the usual x-dot. The notation "$\partial_t$" has long been part of the standard notation in physics.

## 2. Adaptive Control and ADP for Engineering Applications and Intelligent Control

### 2.1. Engineering Applications

In principle, the ideal feedback control system should try to optimize some mix of stability and performance, in the face of three general types of uncertainty:

(1) High-bandwidth random disturbances, which are usually represented as a stochastic process based on random noise [4,5,26,29], but are sometimes represented as bounded disturbances of unknown structure [1];

(2) Drifting values (and occasional abrupt changes) in familiar process parameters such as friction, viscosity and mass, often due to the aging of a plant or changes in general environmental conditions;

(3) Uncertainty about the fundamental structure of the plant or environment (sometimes due to catastrophic events, like a wing being shot off of an airplane), and shifts of parameters in ways that could not be anticipated or simulated even as possibilities at the time when the controller is developed.

This three-fold distinction is difficult to formalize, in mathematical terms, but it has great practical importance. Roughly speaking, the ability to respond to the first type of disturbance or uncertainty may be called "stochastic feedback control;" the second may be called "adaptation;" the third, "learning." The practical tradeoffs here are discussed at some length in the introductory review in [7], and in other papers cited in [7]; here, I will only summarize a few points related to the goals of this particular paper.

"Adaptive control" [1-3] has often been viewed as a tool for addressing the second type of uncertainty -- uncertainty about drifting plant parameters. The most classical designs in adaptive control are intended to control plants $\underline{x}(t)$ governed by the equations:

$$\partial_t \underline{x} = A\underline{x} + B\underline{u}, \qquad (3)$$

or:

$$\underline{x}(t+1) = A\underline{x}(t) + B\underline{u}(t), \qquad (4)$$

where $\underline{u}$ is a vector of controls, where A and B are unknown matrices representing the parameters of the plant, and where $\partial_t$ represents differentiation with respect to time. In the simplest case, the state vector $\underline{x}$ is directly observable. The key idea is to develop control designs which estimate the matrices A and B, explicitly or implicitly, as the process unrolls in real time, so that we can converge "on the fly" to a good control strategy

$$\underline{u}(t) = K(t)\underline{x}(t), \qquad (5)$$

despite our initial ignorance about A and B. In the general case, we assume that we cannot observe $\underline{x}(t)$ directly; instead, we observe some vector $\underline{y}$ governed by:

$$\underline{y}(t) = H\underline{x}(t) \qquad (6)$$

Roughly speaking, this requires the use of a more complex control strategy like:

$$\underline{u}(t) = K_{1,1}\underline{y}(t-1) + K_{1,2}\underline{y}(t-2) + ... + K_{1,k}\underline{y}(t-k) + K_{2,1}\underline{u}(t-1) + ... + K_{2,k}\underline{u}(t-k), \qquad (7)$$

for some integer k. (See [1,p.411].) There exists a huge body of stability theorems for adaptive control, both in the original linear versions [1-3] and in a variety of nonlinear and neural extensions (e.g., [18,19]).

In practice, however, all of these stability theorems require very strong assumptions about the plant or environment to be controlled, as I will discuss in section 3. Ordinary adaptive control and neural adaptive control have often exhibited problems with stability and with slow response to transient disturbances, particularly in real-world plants containing delays and deadzones and reversal phenomena, etc.

Because of these problems, the most powerful approaches available today to cope with type two uncertainty in real engineering applications are:

(1) Adaptive-predictive control [2], in the form which explicitly minimizes tracking error over <u>multiple time periods</u> ahead into the future. This involves a substantial increase in computational complexity, for computations which must be performed in real time.

(2) A two step design process, performed offline before the controller is actually used on the real commercial plant. In the first step, one designs a controller containing very strong feedback loops and/or observers, able to estimate (explicitly or implicitly) the <u>specific</u> plant parameters which are expected to drift. These loops take over the role which the estimates of "A" and "B" would play in an adaptive control approach. In the second step, one exploits <u>prior knowledge</u> about the specific plant, <u>including</u> knowledge about its uncertainties, in order to tune or train this controller. The controller is tuned or trained for optimal performance over multiple time periods, in some kind of offline training process.
Sometimes the optimization is done by using n-period lookahead methods, analogous to model-predictive control [2,8]. Sometimes it is done by dynamic programming approaches.
This approach often has very good response to transient disturbances, because of how it exploits prior knowledge about the plant.

Unfortunately, I personally am not aware of any major success stories based on the first approach; however, Astrom's work in general [2] is well-respected for the many industrial applications it has led to.

Athans and Baras [6] have been very strong spokesmen for the second approach, within the nonlinear robust control community. Athans, in particular, has worked with many aerospace applications, where freezing the dynamics of a controller prior to actual flight can simplify the process of obtaining government approval. Robust control theorists have shown that type one disturbances convert the problem of controller design into a problem of stochastic optimization. (This is true no matter how the disturbances are represented -- either as stochastic noise or as "worst case noise" in the spirit of Tsypkin.) Nonlinear stochastic optimization problems require the use of dynamic programming, rather than a simple
n-step lookahead. In order to find numerical solutions to such stochastic optimization problems, one can simply use learning-based ADP, applied in an <u>offline learning mode</u>.
(As an example, some of the ADP methods to be discussed here involve the use of the classical Galerkin method [27] for solving differential equations, applied to the Hamilton-Jacobi-Bellman equation.) When these methods do converge to a solution of the Hamilton-Jacobi-Bellman equation, one then obtains all the strong guarantees of stability which emerge from this line of research.

In the neural network field, the second approach was first proposed in 1990 [30]. I called it "learning offline to be adaptive online." One can insert the required feedback loops into the controller network, simply by using a Time-Lagged Recurrent Network (TLRN) as a controller [8,ch10; 26,ch.8]. Or one can expand the inputs to the controller, by including the intermediate outputs of TLRNs trained to perform system identification of the plant, in some arrangement; such intermediate outputs serve as neural observers. Learning offline to be adaptive online is the foundation of Ford's "multistreaming approach" [7,31], which underlies the most successful and impressive success of neural networks to date in real-world control applications. Most of the Ford work has used n-step ahead optimization, using backpropagation through time (BTT) to minimize the computational costs; with BTT and special chips, they expect to be able to extend this work to include adaptation on the fly,
so as to upgrade the initial controller developed offline. BTT -- as I first presented it in 1974 [26] -- can be used to reduce the cost of computing exact derivatives through <u>any</u> feedforward differentiable system, not just neural networks. (For systems which are not feedforward, see the review in [32]. See www.nd.com for some tools to implement BTT
and various special cases of TLRN.)

Given this state of the art, many control engineers have questioned the practical value of adaptive control and real-time learning in general. However, continued research on these lines still has practical engineering importance, for at least three reasons:

(1) Even when the stochastic optimization problem is solved offline, improved convergence can be of great practical value. A fully converged solution for J already implies a stability guarantee for the resulting feedback controller; however, the <u>designers</u> of such controllers would benefit from having tools which are more certain to provide these kinds of solution.

(2) Adaptation of parameters or networks "on the fly" -- beyond the scope of substantive prior knowledge or models -- is essentially the only tool available to address type three uncertainty. Like biological brains, we should be able to <u>combine</u> learning on the fly together with application-specific feedback loops (which can also be tuned in real time), in order to get the best of both worlds.

(3) The extensions of adaptive control proposed in this paper may improve stability and performance over time, and thereby enable this class of designs to become more competitive in some practical applications. These methods have inherent advantages over explicit multi-period optimization, such as reduced computational cost and correct treatment of noise in the nonlinear case. It is also possible to combine the two approaches, by using the ADP methods to estimate future costs or payoffs beyond the horizon of the n-step lookahead[15,33].
(Note that this kind of combination is quite different from the "rollout" methods discussed by Bertsekas [24], which use search trees rather than derivative-based methods like BTT.)

## 2.2. True Intelligent Control

Research into intelligent control can benefit more than engineering per se. It can also lead us to develop the kind of mathematical understanding which will be necessary, as a prerequisite, to understanding the kind of intelligence which exists in biological brains.

Many researchers have been overwhelmed by the sheer complexity of biological brains. Particularly in the cerebral cortex of mammals, there is a bewildering array of cells which perform a wide variety of tasks. Many people have despaired of finding any universal principles of information processing at work there, which could be understood in mathematical terms, ala physics. However, studies of mass action in the cerebral cortex -- pioneered by Lashley, Pribram and Freeman, among others -- have demonstrated that cells in one part of the cortex can learn to take over the functions of other cells, when there is a need to do so (and when the required data inputs are not cut off). This suggests that the learning abilities of the higher parts of the brain are relatively uniform and universal, and should permit some kind of global mathematical understanding. (See [9-11,26] for a detailed elaboration of these ideas.) However, the problem lies in developing the right kind of mathematics. Since the overall function of a biological brain is to compute actions or decisions, I have argued that learning-based intelligent control can someday provide the required mathematics. There are many more specific links between ADP designs and the brain, beyond the scope of this paper [9-11]

In general, an understanding of biological intelligence will clearly require a better understanding of real-time learning. Section 5.3 will discuss such issues in more detail.

Nevertheless, the role of real-time learning should not be overstated, even in biology. Since the time of Freud at least, it has been well-known that organisms remember past experiences, and adapt their general models or expectations about their environment based on some combination of present experience and memories. Back in 1977 [25], I proposed a simple learning scheme, based on the more general idea of "syncretism," an interpretation of Freud's vision of the interplay between memory and generalization; syncretism promises substantial practical benefits for neural network adaptation [34; 8,ch.3]. Early successes of "memory-based learning" [35] support this general approach. More recently, McClelland and others have promulgated theories of the interplay between memory and generalization which also embody a form of this idea [36]. My own theories about this phenomenon have predicted that this interplay mainly occurs within the cerebral cortex and the hippocampus, rather than between them, as suggested by McClelland; recent arguments by Pribram, Alkon and others tend to support the former view[36]. But in any case, the psychological evidence for the phenomenon as such is very convincing, in my view. Within the world of artificial intelligence, researchers such as Rosalind Picard of MIT have developed learning methods which also embody this general approach.

In summary, even the brain itself appears to use learning mechanisms which are a kind of hybrid of offline learning and real-time learning. Nevertheless, improved understanding of stability in the real-time case should be helpful for our understanding of the offline aspects as well, as discussed in section 2.1.

## 3. Stable Adaptive Control: Present Limitations and Future Potential

This section will begin by discussing adaptive control in the tradition of Narendra, in the linear case, and then discuss extensions to the nonlinear case and links to some work discussed by Astrom.

### 3.1. Present Limitations in the Tradition of Narendra

In an ideal world, in the linear deterministic case, we would like to possess a universal adaptive controller with the following properties. First, the design and the adaptation rule should tell us how to output the vector **u**(t) at each time t, based only on knowledge about **u**($\tau$) and **y**($\tau$) at previous times $\tau$<t. It should require no knowledge at all about A, B or H, except for the dimensionality of **u** and **y** (and perhaps **x**). It should be guaranteed to send tracking error to zero, asymptotically, at an exponential rate, for all combinations of A, B and H and reference model for which this would be possible with a fixed controller. (The "reference model" is the process -- generally considered linear here -- which outputs the desired observed state **y***(t).) In other words, the ideal adaptive controller should possess the same kinds of strong guarantees of stability which exist for more familiar kinds of control, such as fixed-structure LQG optimal control [4,5].

Adaptive control designs available today fall far short of this ideal. (Again, I am postponing discussion of certain hybrid designs discussed by Astrom[2].) Nevertheless, there is excellent reason to hope that this ideal can in fact be realized, for new true adaptive control designs, as I will discuss.

The bulk of the existing theory for linear adaptive control focuses on the Single-Input Single-Output (SISO) special case -- the case where y and u are scalars. Yet even in that case, stability is normally guaranteed only under three restrictions (in addition to restrictions on the reference model): (1) information must be available which lets us choose the integer "k" in equation 7; (2) the plant must be "minimum phase"; (3) the sign of the high-frequency gain must be known. Narendra and Annaswamy go on to state [1,p.359]: "... it was realized that these assumptions were too restrictive, since in practice they are violated by most plants, even under fairly benign conditions."

The problem with the sign of the gain may be illustrated by a simplified, commonsense example. Suppose that you are a fisherman. You have complete control over the big lake in which you fish. Your "reference model" is simply a certain quota of fish which you must catch every year, in order to satisfy your employers. One month, you discover that did not quite make your quota. The lake is capable of producing enough fish to let you meet the quota, in the long term, but somehow -- due to an unknown disturbance -- the fish population has gone through a minor dieback. Following the usual policy of minimizing tracking error at time t+1 -- the following month -- you simply increase your intensity of fishing that month. Thus you do meet your quota. However, by fishing at a higher level, you reduce the fish population still further -- and dip into the smaller fish which could have provided more growth in the future. Thus to meet your quota in the following month, you have to fish even more intensely. In the end, the entire fish population dies off;
things go nonlinear, and you experience a catastrophic instability. This example leads to a host of further implications. But the key point is this: the policies which reduce tracking error (or maximize utility) in the short term may have undesirable or even catastrophic effects in the more distant future. Knowing the signs of the gains is crucial to avoiding these kinds of problems in adaptive control designs which do not look beyond time t+1.

The "minimum phase" assumption has similar, related pervasive implications, discussed at length by many researchers, including Widrow [37].

Narendra and Annaswamy do discuss methods due to Mudgett and Morse which avoid the need to know the sign of the high-frequency gain, for the SISO case. However, those methods use a very special-purpose trick, the Nussbaum gain, which does not appear to carry over to the general, MIMO case in a serious way. The requirements for prior knowledge in the MIMO case [1, ch.10] are far more complex and demanding than in the SISO case.
After all, in the SISO case, there are only two possibilities for the sign of the high-frequency gain, which is a scalar -- plus or minus. In the MIMO case, there are an infinite number of possible directions or modes of instability, and the requirements are very complex. Narendra has recently found another way to avoid having to know the sign of the gain, by using multiple models [38,39] in the SISO case; again, however, the minimum phase assumption is still required, and a universal MIMO controller developed in this way would be extremely complicated, if possible.

Difficulties with these restrictions probably explain why Lyle Ungar, of the University of Pennsylvania, found unstable results when he tested the most advanced direct, indirect and hybrid direct-indirect adaptive control designs on the bioreactor benchmark test problem given in [22]. There may even be an analogy between the harvesting of cells from a bioreactor and the harvesting of fish from a lake. This same problem was later solved directly and efficiently, both by neural model-predictive-control (based on BTT) and by ADP methods, in papers from Texas Tech [14] and from Ford Motor Company[340.

### 3.2. How To Overcome These Limitations

The previous section leaves us with an obvious, overriding question: **is it actually possible to develop a true adaptive control design which meets the ideal requirements for a universal controller as discussed above?** Beyond that, we face a question which is even more difficult: can we develop nonlinear, stochastic learning controllers which meet all of these requirements when they are applied to linear plants? As it turns out, there is excellent reason to believe that we can do so.

These possibilities are based on the close parallels between a standard Indirect Adaptive Control (IAC) design, illustrated in Figure 1, and the HDP+BAC design illustrated in Figure 2.

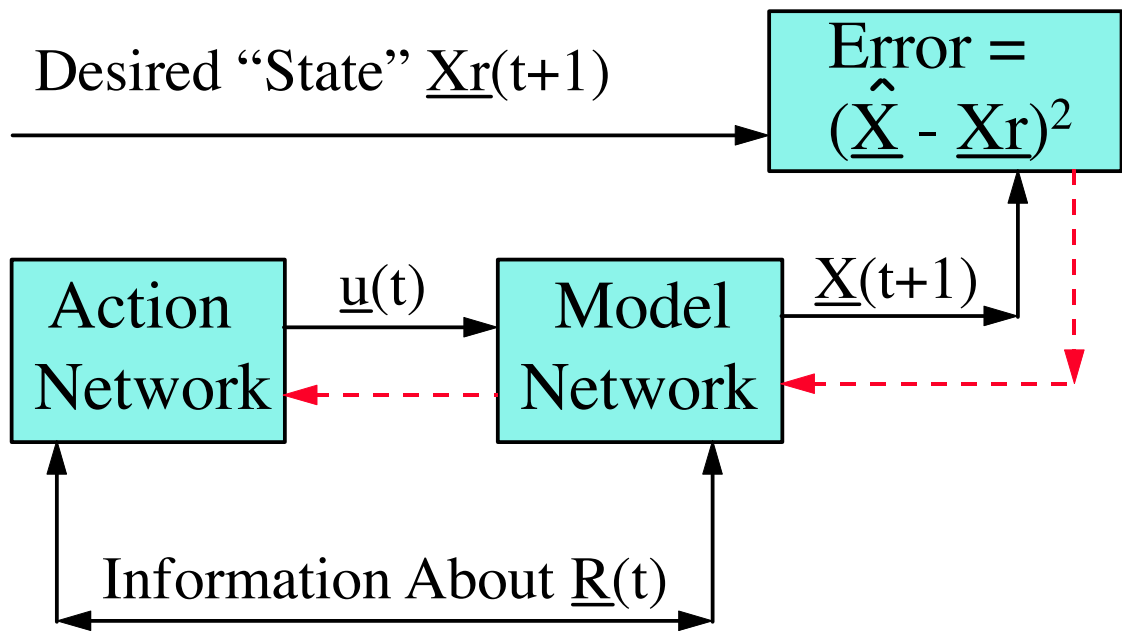

**Figure 1. Illustration of Indirect Adaptive Control (IAC)**

Narendra and Annaswamy [1] discuss both direct and indirect adaptive control designs for the linear case. However, in 1990, Narendra [22,p.135] stated,"... until (unless?) direct control methods are developed, adaptive control of nonlinear dynamical systems has to be carried out using indirect control methods." In 1992, he stated [8,p.168],"... unless further assumptions concerning the input-output characteristics of the plant are made, direct adaptive control is not possible."

Figure 1 is essentially identical to the nonlinear adaptive control design discussed by Narendra in [22,p.135], [8,p.168], [41,p.166] and elsewhere. In the neural network context, I prefer to use the capital letter, **X**, rather than **y**, to indicate the vector of observables. I use **R** to indicate the estimated state vector (or "**r**epresentation of **r**eality," usually based on **r**ecurrent neurons). I use the term "Model network" to describe what he calls the identification network, $N^i$. I use the term "Action network" to describe what he calls the controller network, $N^c$. We both assume that the Model network may be adapted in real time, by an adaptation rule independent from what is used to adapt the Action network; however, on his flowcharts, Narendra adds a few arrows (labeled "$e_i$") to give a hint of how the Model network might be adapted.

Numerous implementations of the design in Figure 1 have been reported in journal articles all over the world. In essence, we only need two key pieces of information here:

(1) How to adapt the Model network. Narendra and I [8,32] have suggested numerous ways of doing this, of varying degrees of robustness. For linear adaptive control, simple least mean squares (LMS) learning may be adequate. (For example, see [1,p.402], [42], and section 7.)

(2) How to adapt the Action network. Narendra proposes that we adapt the weights $W_{ij}$ in the Action network in proportion to the derivatives of tracking error with respect to the weights. This may be written:

$$W_{ij}(t+1) \;=\; W_{ij}(t) \;-\; \alpha \frac{\partial}{\partial W_{ij}} (E(t+1)), \qquad (8)$$

where:
$$E(t+1) = (\underline{e}(t+1))^2 \;=\; (\underline{X}^*(t+1) - \underline{X}(t+1))^2 \qquad (9)$$

is the tracking error at time t+1 and $\alpha$ is some arbitrary (small, positive) learning rate. Narendra asks us to calculate the derivatives in equation 8 by "backpropagating through the Model network." How to do this is described at length both in Narendra's work and -- for a broader class of possible network structures (neural or nonneural) -- in my own [8,26]. The broken arrows in Figures 1 and 2 represent the backwards calculations used to obtain the required derivatives at minimum computational cost. There is also a huge literature on alternative ways to choose the learning rate and on alternative gradient-based learning rules.

Again, however, the design shown in Figure 1 is essentially just a generalization of the linear methods discussed in section 3.1. It has all the same limitations, and perhaps more [41]. Applied to the case of a fully observable linear plant, equation 8 reduces to:

$$K(t+1) \;=\; K(t) \;-\; \alpha B^T \underline{e}(t+1) \underline{x}^T(t), \qquad (10)$$

where the weights in the Action network are now just the matrix K of equation 5. "Backpropagating through a network" (like the Model network, which reduces to equation 4 in the linear deterministic case) is simply a low-cost way to multiply a vector by the transpose of the Jacobian of that network; it reduces costs by exploiting the internal structure of the network, and working backwards.

Figure 2 is very similar to Figure 1, in a mechanical sense, but it has vastly different properties.

In Figure 2, the terms "HDP+BAC" refer to Heuristic Dynamic Programming (HDP) and the Backpropagated Adaptive Critic (BAC). HDP is a technique for adapting the Critic network, which I developed in 1968-1972, to be discussed in section 4. BAC is the method used to adapt the Action network. HDP+BAC is an architecture for optimal control, which is structurally almost identical to IAC.

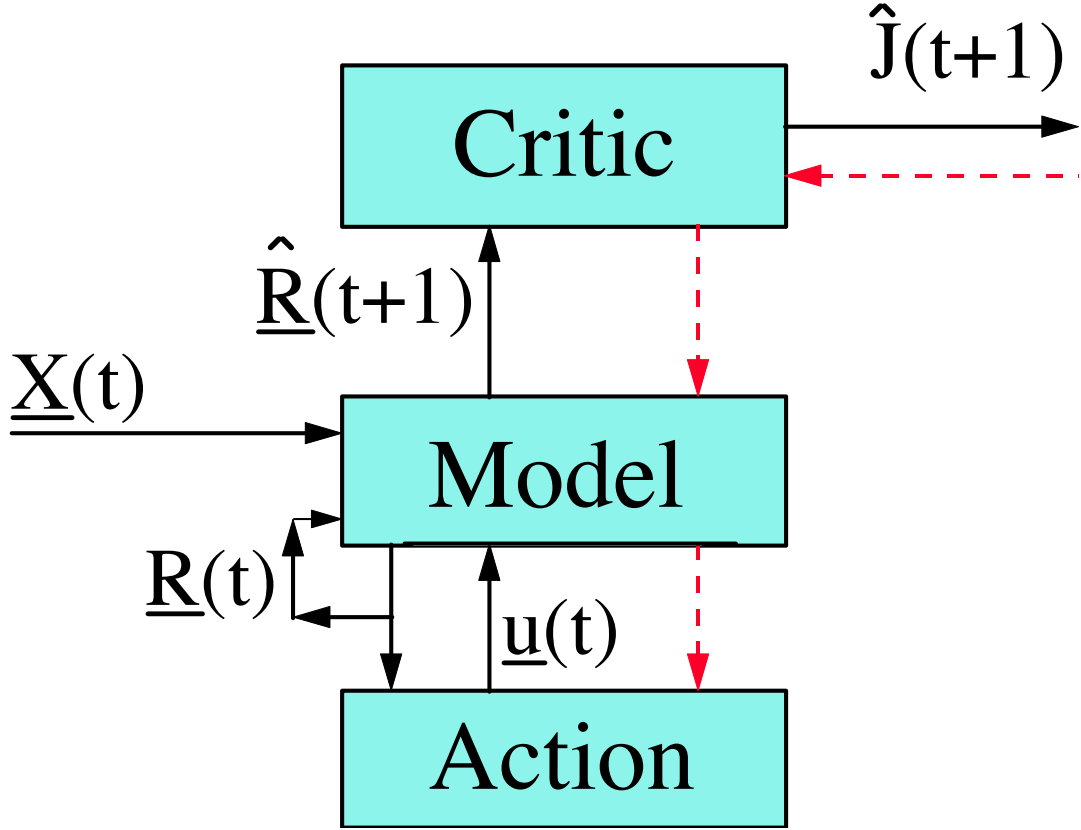

**Figure 2. HDP+BAC**

HDP+BAC can be used in a variety of configurations, involving a combination of real-time learning, offline learning, prior information, and so on. In order to implement a full real-time learning version of HDP+BAC, we need three pieces of information:

(1) How to adapt the Critic network;

(2) How to adapt the Model network;

(3) How to adapt the Action network.

Note that the design here is identical to the IAC design in Figure 1, except that tracking error (E) has been replaced by the Critic network. The Model network can be adapted by using any of the various methods Narendra and I have proposed and used for that task. The Action network can be adapted by replacing equation 8 by:

$$W_{ij}(t+1) = W_{ij}(t) - \alpha \frac{\partial}{\partial W_{ij}}(\hat{J}(t+1)), \qquad (11)$$

where the derivatives are again calculated by backpropagation, in exactly the same way. (To initialize the backpropagation, however, we must first calculate the derivatives of $\hat{J}$ with respect to its inputs, $\underline{R}(t+1)$; these derivatives can be calculated quite easily by backpropagating through the Critic network. In the deterministic case, the distinction between **R** and **R̂** is not so important as it is in the stochastic case [8].) Strictly speaking, Figure 2 represents a special case of HDP+BAC applicable to the tracking problem; see section 4.3.2 for a slight generalization.

In other words, to implement Figure 2, one can use the same computer code used to implement Figure 1, except that "E" is replaced by "$\hat{J}$."

HDP attempts to adapt the Critic network in such a way that its output, $\hat{J}$, converges to a good approximate solution to the Bellman equation of dynamic programming. This equation may be written fairly generally [8] as:

$$J(\underline{R}(t)) = \underset{\underline{u}(t)}{\text{Max}} \ \{U(\underline{R}(t), \underline{u}(t)) + (1/(1+r)) <J(\underline{R}(t+1))>\} - U_0, \quad (12)$$

where U is the utility or cost function which we want to maximize or minimize in the long term, where r is an interest rate parameter used to discount the value of future utility, where angle brackets <> indicate expectation value, and where $U_0$ is a parameter introduced by Howard [28] to extend dynamic programming to the case of an infinite time horizon with r=0. (Of course, for a minimization task, we replace "Max" in equation 12 by "Min.") When we apply this method to pure tracking problems, as in classical adaptive control, we may simply choose U to be the tracking error E, and treat the reference model as a fixed augmentation of the Model network.

**There are three main reasons to expect that something like HDP+BAC could yield a universal stable adaptive controller in the linear case as discussed above:**

(1) In linear quadratic optimization problems, it is well known that the solution to Bellman's equation is simply quadratic in the state of the system; therefore, by choosing the "Critic network" to be a general quadratic function of the state, we can be sure that some value of the weights will yield an exact solution of the Bellman equation. Of course, we must check that we really fulfill this condition as we carry out this research.

(2) If a deterministic plant of this sort is controllable at all, then there will always exist some control law which sends the tracking error to zero, at an exponential rate. This guarantees that there exists a strategy or policy of action (control rule) for which the sum of future cost will be finite, even with $r=U_0=0$. Furthermore, dynamic programming always yields the strategy of action of minimum total future costs, when a minimum exists; the previous sentence implies that a minimum does exist, and we may be sure that the dynamic programming solution will do as well or better.

(3) Landelius [43] has already proven whole-system stability results for a particular implementation of HDP+BAC and for other methods to be discussed in section 4. Unfortunately, his new ways of implementing these methods -- while very reasonable from a practical point of view -- are technically not parallel distributed systems.They are not true adaptive control designs in the spirit of Narendra, because they do not meet the tight constraints on computational cost and structure which are required for neural networks. Nevertheless, they are an important achievement. They go far beyond the earlier, similar work of Bradtke [44] in addressing total-system stability issues. They provide a strong indication that this line of research can be successful.

In summary, if a parallel distributed version of HDP does in fact make the Critic converge to the dynamic programming solution, then we would expect some corresponding form of HDP+BAC to provide us with a universal adaptive controller for the linear case.

Note that such a structure would be a true adaptive control design, in the spirit of Narendra, because <u>all</u> of the concurrent adaptation laws (including the versions of HDP to be discussed in section 4) are low-cost, single-pass adaptation laws, consistent with the sort of local, parallel computing found in the brain -- a style of computing which also makes it possible to build very high throughput dedicated chips.

Even for the fully observable linear deterministic case, it will take substantial effort to prove <u>total system stability</u> for this class of design, accounting for the coupled dynamics of all four parts of the overall system -- the plant, the Action network, the Model network and the Critic. It took many years before Narendra and others [1-3] proved restrictive theorems about total system stability even for the SISO state, for systems with only three major components (plant, Model, Action).

The first, most critical step in this process is to prove stability for the Critic adaptation as such, holding the rest as fixed, for a Critic adaptation rule which is also applicable to the nonlinear, stochastic case. This paper began as an effort to take that first step. However, none of the existing Critic adaptation rules were able to pass that test, for the definition of stability to be discussed in section 7. The main contribution of this paper is to provide new adaptation rules which do.

### 3.3. Adaptive Control in the Traditions of Astrom and of Liapunov

Researchers in control theory will undoubtedly find many links between the approach proposed here and other strands of research. Of these various links, the most important that I know about are the links to the ideas of Astrom [2] and links to a form of Liapunov-based control, where a fixed Liapunov function developed by the designer takes the place of the
adapted Critic network in Figure 2. This section will discuss these two sets of links.

### 3.3.1. Links to the Astrom Tradition

Karl Astrom has perhaps been the world's number one leader in stimulating real-world applications of adaptive control. He has done this in part by developing sophisticated hybrid designs, in order to meet the demanding requirements of these applications. For example, he played a pioneering role in building a link between linear/quadratic optimization and adaptive control. In [8,ch.2], he and McAvoy give a broad overview of efforts by themselves and others to address the general subject of intelligent control. The latest edition of his classic
text on adaptive control [45] contains many references to new work in these directions.

The original, classic text by Astrom and Wittenmark [2] discusses three types of adaptive control which overcome many of the restrictive assumptions discussed by Narendra:
(1) multiperiod adaptive-predictive control; (2) adaptive pole placement; (3) linear-quadratic Self-Tuning Regulators (STRs). (In fact, on June 18, 1992, Astrom was kind enough to send me an email on the minimum phase problem; he pointed towards the adaptive pole placement and LQG/STR approach as the most important ways to overcome this problem.)

Multiperiod adaptive-predictive control was already discussed in section 2.1. The future plans by Ford Motor Company to consider using multiperiod optimization in real time, along with real-time system identification, could be viewed as a possible future application of this approach. There are many other links to the most successful applications of artificial neural networks [7]. Methods of this family are of great practical utility, but they are not "true" adaptive control designs as discussed in section 3.2. They are also not plausible as a theory of how the brain actually implements intelligent control.

Adaptive pole placement appears to be even further away from being a "true" adaptive control design as discussed in section 3.2. Also, it is hard to imagine a nonlinear stochastic generalization of that design. The discussion in [2] describes the method as a SISO method.

On the other hand, the discussion of linear-quadratic STRs in section 5.5 of [2] has many close links to the approach proposed here. One version, based on spectral factorization, does not appear relevant to my goals here. But the other version -- Indirect STR based on Riccati equation -- is extremely close. This version is called Algorithm 5.7 in [2], and Algorithm 4.4 in [45]; the new edition of the text is almost identical to the old version in this section. The matrix $S(t)$ in Astrom's equation 5.47 is essentially identical to the matrix of Critic weights C discussed in this paper. The methods used by Astrom and others to update the estimate of $S(t)$ could be viewed as specialized Critic adaptation methods, for use in the linear/quadratic case. Unfortunately, the most reliable update methods -- based on solving a Riccati equation or performing a spectral factorization -- are not true adaptive control methods as defined above. Astrom hints that there have been serious stability problems when other update methods have been attempted; however, he also hints that D.W.Clarke of Oxford and M.Karny of Prague have been more central to this work in recent years. On a quick examination of their web pages, I have not found material which would change the conclusions of this paper. However, Karny has certainly made significant contributions to the larger area of intelligent systems [46].

Landelius' method of implementing HDP was motivated in part by the most recent work in this literature [47], which he also cites and discusses.

### 3.3.2. Control Based on Liapunov Functions

Liapunov stability theory in general has influenced huge sections of control theory, physics, and many other disciplines. More narrowly, within the disciplines of control theory and robotics, many researchers have tried to stabilize complex systems by first deriving Liapunov functions for those systems. In some cases, the Liapunov functions have been derived analytically by solving the multiperiod optimization problem in an analytic fashion. The details of this stream of research go far beyond the scope of this paper, and beyond the knowledge of its author.

After one has derived such an application-specific Liapunov function, one can then use a design exactly like Figures 1 and 2, except that the Liapunov function replaces the square tracking error or J. Theoretically, by replacing the square tracking error with some other prespecified error measure, one arrives at a whole new class of stability properties, and a whole new set of restrictions on the plant.

The Liapunov function approach has been particularly useful in the field of robotics [48], where many robot arms obey nonlinear but rigid dynamics, which have been analyzed in depth. Robert Sanner of the University of Maryland and John Doyle of CalTech are often mentioned in discussions about the successes of this approach. Nevertheless, none of these analytically derived fixed Liapunov functions can be regarded as a universal adaptive controller, as discussed above, any more than IAC itself can.

From a practical point of view, it becomes more and more difficult to derive such Liapunov functions analytically, as one tries to control more and more complex nonlinear systems, such as elastic, light-weight and flexible robot arms. (See [7] for an explanation of some of Hirzinger's success in working with such robots. Some of the unpublished work of Fukuda has been even more useful.) The difficulties here are analogous to the difficulty of trying to solve simple algebraic equations analytically. As one progresses from quadratic equations, to cubic equations, to quartic equations, to sixth order equations, and so on, one eventually reaches a point where closed-form analytic methods simply can't give you the solution. At some point, one has to use computer-based numerical methods instead of analytical methods.

<u>Critic adaptation methods are a way to derive Liapunov functions in a computer-based numerical fashion</u>. The J function of dynamic programming is guaranteed to be a Liapunov function, at least for deterministic systems. This connection points towards another new direction for possible research, beyond the scope of this paper [49]. One way of implementing this connection would be to use analytically-derived Liapunov functions as <u>initial values</u> for Critic networks; for example, a Liapunov function developed for rigid-body robotics could be tuned further, adaptively, using a method like HDP, in order to derive a more accurate Liapunov function for a somewhat flexible robot. By successive tuning, one could derive Critics suitable for the control of much more flexible robots. Section 5 will discuss one other possible way to exploit these connections.

## **4. Learning-Based Approximate Dynamic Programming (ADP): Standard Methods for Adapting Critics and New Variations**

### **4.1. Introduction**

Hundreds of papers have been published by now on various forms of approximate dynamic programming (ADP), adaptive critics and/or reinforcement learning. However, only three of the established, standard methods for adapting Critic networks are directly relevant to the goals of this paper: (1) Heuristic Dynamic Programming (HDP); (2) Dual Heuristic Programming (DHP); and (3) Globalized DHP (GDHP). The classic "Temporal Difference" (TD) method is essentially a special case of HDP. The recent "two-sample" method [24,50,51] is also relevant, but will not be analyzed until section 9.

Section 4.2 will give an oral history of the fundamental methods available in this field, their origins and motivations. It will describe why the remainder of the paper will focus on variations of HDP, DHP and GDHP, up until sections 8 and 9. Section 4.3 will define what these three methods are, and also give Galerkinized and continuous-time versions of them, for the general nonlinear stochastic case. Finally, section 4.4 will derive the special cases of these methods for observable linear-quadratic systems. Later sections of this paper will provide other new methods for Critic adaptation.

Section 3.2 explained why our initial efforts here have focused on demonstrating stability for methods of adapting Critic networks as such -- not for entire adaptive critic control systems. Therefore, this paper will not provide a complete review of the larger question of how to adapt a complete ADP control system. For the concurrent adaptation of Model networks, Action networks and Critic networks, and for practical experience, see [8,14,15,33,52-56].

The only existing, working control designs which meet my criteria/definition for "Model-Based Adaptive Critics" (MBAC) or "brain-like intelligent control" [33] are those based on Figure 2, or the equivalent of Figure 2 for DHP, GDHP and their variations. MBAC has been successfully implemented by at least three companies (Ford Motor Co., Accurate Automation Corp., Scientific Cybernetics Inc.), four professors working with graduate students (Wunsch, S.Balakrishnan, Lendaris, W.Tang) and three other individuals (Jameson, Landelius and Otwell). Since the first early implementations in 1993, MBAC has outperformed other modern control and neurocontrol methods in a variety of difficult simulated problems, ranging from missile interception (Balakrishnan) through to preventing cars from skidding when driving over unexpected patches of ice (Lendaris). The one physical implementation published to date was also highly successful (Wunsch's solution of Zadeh's "fuzzy ball and beam" challenge). Balakrishnan has said that he will report on a larger physical implementation (a cantilever plate) at the forthcoming American Control Conference. The one alternative method which still seems competitive with MBAC, in terms of performance, in these tests, is neural model-predictive control based on BTT, discussed in section 2.1. Again, however, the remainder of this paper will focus on the problem of adapting Critics <u>as such</u>; therefore, it will not discuss the other aspects of this work.

**4.2. Basic Methods for Adapting Critics: Early Origins, Motivation and Selection**

Roughly speaking, the adaptive critic field of research is a single new research field (albeit still divided into factions) which emerged around 1988-1990 through the unification of several previously separate strands of research. The origins of the field up to that time are summarized in Figure 3. Within the field itself, the terms "reinforcement learning,"
"adaptive critics," "approximate dynamic programming" and "neurodynamic programming" are normally viewed as approximate synonyms. Nevertheless, the choice of terms also reflects different goals for research within the field and different patterns of interest in related research outside of the field. Please note that Figure 3 only represents the flow of key ideas in this topic prior to 1988-1990; it does not represent patterns of personal association, ideas in other areas, or important recent ideas which will be mentioned
later on in this section.

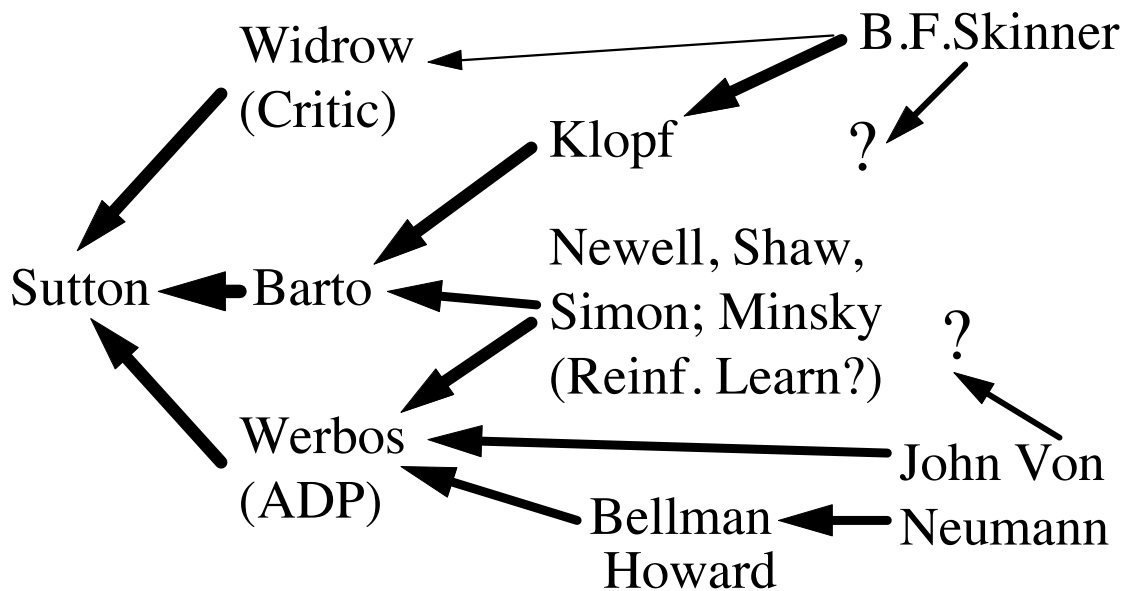

**Figure 3. Early Origins of the Adaptive Critic Field**

The psychologist B.F. Skinner is well-known for the idea that rewards and punishments ("primary reinforcement signals") determine the behavior of animals including humans.
Many of the ADP designs used today in engineering or computer science are also being used as empirical models of animal behavior. Harry Klopf was a major pioneer of that tradition; Klopf, in turn, was responsible for recruiting and funding Barto to explore this strand of research. (Some of Klopf's critics accused him of using Air Force research funds for "self-promotion," for the promotion of his own ideas; nevertheless, the research which he funded was clearly of critical importance to the early development of this field.) Barto has continued this tradition through ongoing collaborations with psychologists and neuroscientists.
Widrow never pursued these kinds of connections, but his seminal 1973 paper on adaptive critics [23] was clearly influenced by Skinner's notion of reward and punishment. Skinner's notion of a "secondary reinforcement system" can be viewed as one way of talking about a Critic network.

Von Neumann and Morgenstern [57] invented the concept of cardinal utility function, which underlies the ADP approach. This work also made possible many new directions in economic theory, ranging from game theory to decision analysis and Bayesian utilitarianism. Working with Richard Bellman, Von Neumann also helped to inspire the development of dynamic programming. In criticizing the neuron models of McCulloch and Pitts, Von Neumann [58,p.451] made two major points: (1) that neuronal signals may be more accurately modeled as continuous variables rather than binary signals in many cases; (2) that study of mechanisms like learnng and memory are more important than efforts to understand how particular static, learned functions can be represented in fixed neural networks. The successful revival of the neural network field in the past decade was based, in large part, on researchers finally embracing these two crucial insights.

On a philosophical level, Skinner and Von Neumann were extremely far apart. Skinner's words about molding and controlling human beings certainly gave encouragement (even if unintentionally) to a variety of ideologies in the spirit of fascism and Communism. The revival of the neural network field in the 1980's was motivated in part by a rejection of Skinner's approach and the resulting search for new paradigms in psychology[59]. Von Neumann, on the other hand, encouraged a higher level of respect for human intelligence, autonomous human decision-making and human potential, from the very beginning. Despite these differences, the most basic views and insights of both men have been fully incorporated into the ongoing research in this field. Different researchers have different attitudes about the deeper philosophical implications [12], even as they use the same mathematics.

The concept of reinforcement learning as a pathway to the construction of intelligent systems has often been credited to the great pioneers of artificial intelligence (AI) -- Newell, Shaw and Simon -- and to Marvin Minsky [60]. They proposed the development of machines which learn over time to maximize some measure of reward or reinforcement. They proposed that such a machine, simply by learning through experience, could gradually develop higher-order intelligence as a kind of emergent phenomenon. The earliest attempts to implement this idea were based more on brute-force stochastic search rather than optimization theory, and the results were somewhat disappointing. Samuels' classic checkers-playing program [60] has been interpreted, in retrospect, as a kind of adaptive critic system; his adaptive "static position evaluator" served as a kind of Critic. More recently, Tesauro's master-class backgammon program [24,55] has clearly demonstrated that reinforcement learning systems can generate intelligent behavior.

In 1968, I published a paper [20] which argued that reinforcement learning could be used as a foundation for understanding intelligence in the brain. I formulated the concept of reinforcement learning in the modern way (illustrated in Figure 4). In this concept, a system <u>learns</u> a strategy of action which tries to maximize the long-term future expected value of <u>utility</u>, following the concepts of Von Neumann. For the first time, I pointed out that one could build a machine to perform reinforcement learning simply by trying to approximate dynamic programming as formulated by Howard [28]. This paper also described much of the intuition behind backpropagation and links to Freud, etc.

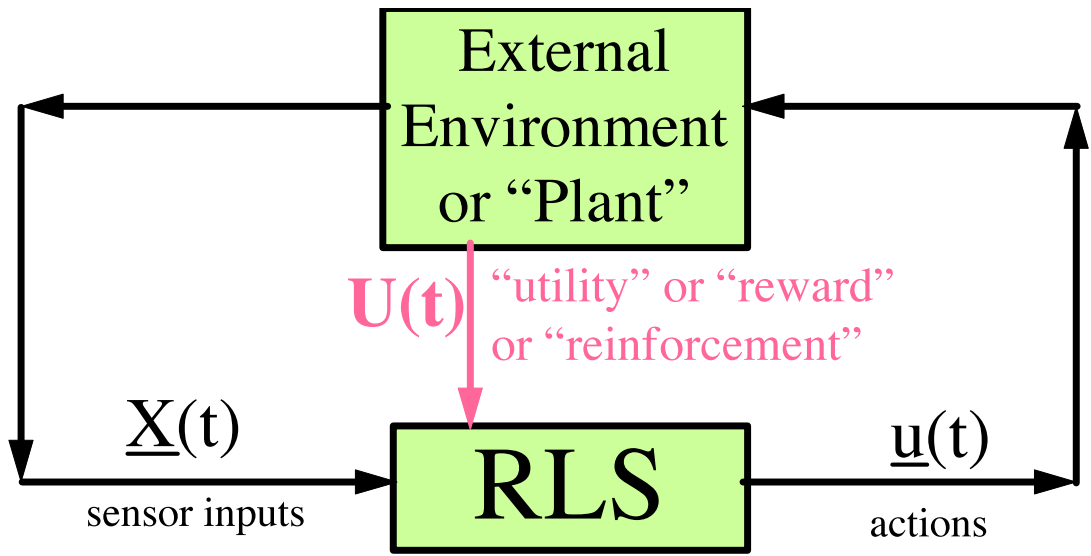

**Figure 4. Reinforcement Learning Systems**

However, the procedures which I then suggested for adapting a Critic (which I then called an "Emotional System") have not been followed up on, by myself or others They were based on an effort to use older approaches, such as feature extraction and linear, more Hebbian-style learning rather than backpropagation.

In 1972, in my Harvard PhD thesis proposal, I included a flowchart virtually identical to Figure 2, with a detailed discussion of how to adapt the components of the system. However, my committee insisted that I focus the thesis on a rigorous, generalized formulation of backpropagation as such, along with a demonstration of its effectiveness in system identification and political forecasting and a minimal discussion of neural networks [26,61].
(The thesis is reprinted in its entirety in [26].) In any case, I was concerned that HDP would have certain difficulties in learning speed, as one scales up to larger problems [8,22]. From 1977 to 1981, I published the ideas of HDP and of two more sophisticated methods -- DHP and GDHP -- intended to overcome these scaling problems [25,62,63]. In particular, [63] discussed the idea of using a neural network trained by backpropagation in order to approximate the J function of dynamic programming. Section 4.3 will discuss these methods in detail.

In the meantime, on a totally independent basis, Bernie Widrow [23] published the first working, successful implementation of an adaptive critic, using a neural network as a Critic network. His 1973 paper is the original source of the term "Critic." However, the Critic adaptation method which he used has not been used elsewhere. Also, it was not a true real-time learning method.

In 1983, Barto, Sutton and Anderson [21] published a classic paper which became extremely famous in the 1990's. They implemented an adaptive critic system consisting of two adaptive elements or neurons. One element was a Critic, trained by a method which Sutton calls a Temporal Difference (TD) method. This method was developed on the basis of intuitive arguments about how to improve upon Widrow's algorithm. The method used in that paper was a very narrow special case of HDP. Later, in a famous 1988 paper [64], Sutton [64] expanded the concept of temporal difference methods to include "TD($\lambda$)" for $\lambda$ in the interval [0,1]. TD(1) essentially reproduces the old Widrow method, and TD(0) a generalization of the method in [21]. Very few papers actually use TD($\lambda$) for $\lambda$ other than zero (see 15,24,55). The other element of the system in [21] was a kind of Action network, trained by an associative reward/punishment algorithm called Arp, which is not relevant to the goals of this paper, because it is not a method for adapting Critics as such.

In January 1987 [65], I published a paper on ADP, GDHP, potential engineering applications and links to neuroscience, once again emphasizing the connection between reinforcement learning and dynamic programming. This paper was noticed by Sutton. It led to a number of further discussions, which led to major efforts by Barto and Sutton to follow up on this new connection to dynamic programming. Barto encouraged the early efforts in this area by Bertsekas [24], who, along with Tsitsiklis, has substantially enlarged the concept of TD. The theory behind TD has been expanded considerably by these and other efforts; however, the method itself remains a special case or subset (proper or otherwise) of HDP.

In the 1987 discussions with Sutton, I also stressed the importance of offline simulation or "dreaming" as a way of improving the performance of ADP systems (as previously discussed in [65]). This was the basis of the "Dyna" architecture first presented by Sutton in [22]. As section 2.2 has discussed, this kind of offline learning is extremely important; however, this paper will focus on the real-time case. Current evidence from psychology is consistent with this as a partial interpretation of dreams in human beings as well [66].

The discussions of 1987 also contributed to the arrangement of an NSF workshop on neurocontrol, held in New Hampshire in 1988, which resulted in [22]. That workshop and book were in my view responsible for brining together neurocontrol and adaptive critics as organized fields of research. Figure 3.1 of that book illustrated the modern concept of reinforcement learning in a more picturesque way, equivalent to Figure 4 above.

Earlier forms of "reinforcement learning" which did not address the mathematical problem of maximizing utility (or minimizing cost) over time [67] would not help us in carrying through the strategy given in section 3.2.

More recently, another family of methods has been developed which I call "Action-Dependent Adaptive Critics" (ADAC). This generally includes Q-learning, ADHDP, ADDHP and ADGDHP [8,14], all of which are closely related. In fact, many of the "new" designs for "extended" or "modified" or "policy" Q-learning are actually implementations of ADHDP, which was reviewed at length in 1992 [14]. That book also reported a successful implementation at McDonnell-Douglas in the manufacturing of composite parts. It reported a successful simulated application to the control of damaged aircraft, which later led to a large related program at NASA Ames. All of these methods use Critic networks which input not only $\underline{\mathbf{R}}$(t) but also $\underline{\mathbf{u}}$(t). Instead of approximating the J function of dynamic programming, they approximate a related function. But for purposes of this paper, I am focusing on the problem of Critic adaptation where the controller or Action network is held fixed. Under these conditions, the action-dependent methods for Critic adaptation reduce to their older, action-independent equivalents.

Another family of reinforcement learning methods called Alopex has also become popular in certain circles. However, the most effective version of Alopex which I have seen is due to Tzanakou of Rutgers. In that version, the Critic network is an action-dependent system, adapted by a procedure quite similar to ADHDP or Q-learning. Most of Klopf's final designs, in collaboration with Leemon Baird, have the same property. As with Q-learning and ADHDP, they do not appear relevant to the goals of this paper. However, the two-sample method [50] discussed in section 9 is a significant exception.

In certain special cases, the same mathematics which underlies Figure 2 can be implemented without a Model network as such. For example, in a plant with a single state variable, it is enough to know the <u>sign</u> of (∂R(t+1)/∂u(t)) in order to compute the derivatives of $\hat{J}$(t+1) with respect to the weights, to within a scalar factor which effectively just modifies the learning rate α. This kind of effect has permitted the development of architectures similar to Figure 2, but simpler, with comparable performance in certain special cases; for example, see some of the work by Lewis [68] and by Berenji [69]. This does not affect the adaptation of the Critic as such, which is the main topic of this section.

More recently, I have developed several new families of adaptive critic designs, most notably the Error Critics and a multiple-model hierarchical decision system, both motivated by recent findings from neuroscience [8-10,13,32,46]. The Error Critics were intended to met certain important but secondary requirements; I do not yet see how they could be used to meet the primary goals of this paper. (However, the paper by Prokhorov in [50], linked to [15], may suggest possibilities.) The hierarchical decision system is a higher-level learning design, which essentially requires ordinary Critic networks as subsystems; it does not address the immediate goals of this paper, though it <u>is</u> relevant to the longer-term strategy discussed in section 5.

In summary, HDP, DHP and GDHP are the three Critic adaptation methods relevant to the immediate goals of this paper. TD, as a subset (proper or improper) of HDP, is automatically included in this analysis. The two-sample method will be discussed in section 9.

### **4.3. Specification of HDP, DHP and GDHP in the Nonlinear/Neural Case**

### **4.3.1. General Definitions of the Methods in the Discrete Time Case**

HDP and GDHP are methods for adapting either neural networks or other parametrized systems $\hat{J}(\underline{R}(t), W)$ -- with an array W of parameters or weights -- so as to approximate the J function of dynamic programming. The J function of dynamic programming is the function J which satisfies the extended Bellman equation (equation 12).

In Figure 2, the Action network or controller implements a mathematical function $\underline{u}(\underline{R}(t))$. In the language of dynamic programming [24], this function is called a "policy" (π) or a strategy of action. The key theorem of dynamic programming is that the choice of actions $\underline{u}$ implied by equation 12 represents the best policy of action; in other words, for the case where $U_0=0$, it yields the policy π such that:

$$J(\underline{R}(t)) = \underset{\pi}{\text{Max}} \ J^\pi(\underline{R}(t)), \qquad (13)$$

where $J^\pi$ is the expected value under policy π:

$$< \sum_{k=0}^{\infty} U(\underline{R}(t+k), \underline{u}(t+k)) / (1+r)^k > \qquad (14)$$

For a <u>fixed</u> policy of action π, simple algebraic manipulations [24,44] yield an equation parallel to equation 12:

$$J^\pi(\underline{R}(t)) = U(\underline{R}(t), \underline{u}^\pi(t)) + (1/(1+r)) <J^\pi(\underline{R}(t+1))> - U_0^\pi, \quad (15)$$

where $U_{0^\pi}$ is the expected value of U across all future times under policy π. In finite-horizon problems with r=0, the best policy is the one which maximizes $U_{0^\pi}$; in the absence of "crossroads phenomena," it can be obtained by choosing **u**(t) as in equation 12 for the maximal $U_{0^\pi}$. [28].

DHP attempts to adapt a network $\hat{\underline{\lambda}}(\underline{R}(t), W)$ so as to approximate the vector:

$$\underline{\lambda}(t) = \nabla_{\underline{R}} J(\underline{R}(t)), \qquad (16)$$

i.e.,

$$\lambda_i(t) = \frac{\partial}{\partial R_i} J(\underline{R}(t)) \qquad (17)$$

These $\lambda_i$ variables correspond to Lagrange multipliers [4,5] in deterministic optimal control, and to the "costate variables" in Pontryagin's approach. In economics, such Lagrange multipliers are associated with price signals, or, more precisely, with "shadow prices;" thus from an economist's viewpoint, DHP may be regarded as a distributed, adaptive system for calculating prices under conditions of uncertainty.

The recurrence relation for $\underline{\lambda}$ may be derived simply by differentiating equation 12; in the simplest, most direct representation, this yields [8, ch.13]:

$$\begin{aligned}
\lambda_i(\underline{R}(t)) &= \frac{\partial J(\underline{R}(t))}{\partial R_i(t)} = \frac{\partial U(\underline{R}(t),\underline{u}(\underline{R}(t)))}{\partial R_i(t)} + \frac{1}{1+r} < \frac{\partial J(\underline{R}(t+1))}{\partial R_i(t)} > \\
&= \frac{\partial U(\underline{R}(t),\underline{u}(\underline{R}(t)))}{\partial R_i(t)} + \sum_j \frac{\partial U(\underline{R},\underline{u})}{\partial u_j} \cdot \frac{\partial u_j(\underline{R}(t))}{\partial R_i(t)} \\
&\quad + \frac{1}{1+r}\left\{\sum_j < \frac{\partial J(\underline{R}(t+1))}{\partial R_j(t+1)} \cdot \frac{\partial R_j(t+1)}{\partial R_i(t)} > + \sum_{j,k} < \frac{\partial J(\underline{R}(t+1))}{\partial R_j(t+1)} \cdot \frac{\partial R_j(t+1)}{\partial u_k(t)} \cdot \frac{\partial u_k(t)}{\partial R_i(t)} > \right\} \\
&= \frac{\partial U(t)}{\partial R_i(t)} + \sum_j \frac{\partial U(t)}{\partial u_j} \cdot \frac{\partial u_j(t)}{\partial R_i(t)} + \frac{1}{1+r}\left\{\sum_j \lambda_j(t+1)\left(\frac{\partial R_j(t+1)}{\partial R_i(t)} + \sum_k \frac{\partial R_j(t+1)}{\partial u_k(t)} \cdot \frac{\partial u_k}{\partial R_j}\right)\right\}
\end{aligned} \qquad (18)$$

Unfortunately, equation 18 -- unlike equation 12 -- no longer includes the recipe for choosing **u**. However, we still require (as in equation 12) that **u** should be chosen so as to maximize (or minimize) U(t)+<J(t+1)>/(1+r). For differentiable systems, this implies:

$$\begin{aligned}
0 &= \frac{\partial}{\partial u_i}(U(t) + \frac{1}{1+r} < J(t+1) >) \\
&= \frac{\partial U(t)}{\partial u_i(t)} + \frac{1}{1+r}\sum_j < \frac{\partial J(\underline{R}(t+1))}{\partial R_j(t+1)} \cdot \frac{\partial R_j(t+1)}{\partial u_i(t)} > \\
&= \frac{\partial U(t)}{\partial u_i(t)} + \frac{1}{1+r} < \lambda_j(t+1) \cdot \frac{\partial R_j(t+1)}{\partial u_i(t)} >
\end{aligned} \qquad (19)$$

Equation 18 is a stochastic version of the classical Pontryagin principle. In recent years, the mathematics community has shown interest in other, more formal stochastic versions of the Pontryagin principle [70]. For any particular policy π, the differentiation of equation 15 again yields equation 18, but without the additional requirement of equation 19.

**The remainder of this paper will consider the special cases of HDP and DHP** in which the Critic networks ($\hat{J}$ or $\hat{\underline{\lambda}}$) are adapted by gradient-based error minimization, as in classical backpropagation [26]. However, in the neural network community, many researchers prefer to use different types of adaptation, in order to perform basic tasks in supervised learning. Thus in [8,17] I specified in detail exactly how to implement HDP in the general case, where the user is allowed to choose whatever supervised learning system he or she prefers. In [8], I did likewise for DHP. This is the reason why Landelius [43] was able to use an alternative supervised learning method, somewhat analogous to Recursive Least Squares (RLS), in his implementations of HDP, DHP, etc. (RLS is not a parallel distributed processing method, according to the strict standards of neural network research, but it comes close.) Unfortunately, in equation 10 of [8, ch. 13], there is a significant typographical error; the left-hand side should read "$\underline{\lambda}$*(t)".) The reader should also be warned that the pseudocode in [22] contains typographical errors so severe that it is almost useless. GDHP inherently requires a gradient-based learning approach.

In practice, the restriction to gradient-based learning is not as serious as it may seem at first, because many popular learning algorithms can be expressed in that form. For further examples and discussions of how to implement these methods in the general case, see the prior literature discussed in section 4.1. Certain kinds of second-order information can also be used in true gradient-based parallel processing designs, as will be discussed later in this paper. Landelius, in correspondence, has suggested that his novel ways of implementing DHP and ADDHP might provide an alternative route to meeting the goals of this paper, in terms of low-cost universal adaptive control.

To adapt the Action networks, in the general case, one mainly uses the derivatives of U(t)+<J(t+1)> with respect to the weights in the Action network. One can calculate those derivatives by first calculating the derivatives with respect to **u**(t) (as in equation 19, implemented by backpropagation), and then using backpropagation again to propagate them through the Action network. For HDP and GDHP, this is exactly like the discussion of section 3.2, except that an extra term is added for the derivatives of U(**R**(t), **u**(t)) with respect to **u**(t). Because of my emphasis here on the classical tracking problem, the remainder of this paper will consider only the special case where U(t) is expressed simply as U(**R**(t)), so that this additional term drops out.

Even though equation 19 calls for the use of an expectation value, one can achieve the same change of weights, on average, by simply adapting the weights in response to the currently observed derivatives, as in the usual LMS learning and statistical sampling.

In HDP and GDHP, it is possible to supplement this training of the Action network, to include some stochastic search (particularly in the offline mode) to help keep the system out of local minima [46]. Barto, Thrun and others have discussed other approaches to adding noise to the Action network or controller [8,21].

**4.3.2. HDP, DHP and GDHP and Galerkinized Versions For This Paper (Discrete Time)**

The remainder of this paper will assess the stability and accuracy of Critic adaptation in situations where the Action network or policy is held fixed, where the Model network is also held fixed, where U(t) does not depend on **u**(t), and where Critic adaptation is based on the obvious gradient-based versions of the Critic adaptation methods. The reason for this was discussed at length above. In addition, as discussed in section 3.2, I will assume a completely observable plant, where **R**(t) and **X**(t) are the same. For simplicity, I will assume that $U_0=0$ or that $U_0$ is known apriori.

In this special case, there really is no Action network, in effect. We may define the augmented Model network as the combination of the Model network proper and the Action network, treated as one large network. We may write the augmented Model network as:

$$\underline{R}(t+1) = \underline{f}(\underline{R}(t), \underline{w}(t)), \tag{20}$$

where $\underline{w}(t)$ refers to a vector of random variables, as in [39]. (Statisticians more often use the notation $\underline{e}(t)$ or $\underline{u}(t)$ to refer to a vector of random numbers; however, this notation has led to severe confusion in parts of the control theory community, especially involving the definition of "ARMA" processes, which are explained at length in [26,29].)

In real-time adaptation in HDP, the Critic is adapted as follows at each time t:

(1) First obtain the vector $\underline{R}(t)$, simply by observing (or estimating) the current state of the plant.

(2) Next obtain a sample of $\underline{R}(t+1)$, either by <u>simulating</u> equation 20 or by <u>waiting</u> until time t+1 and obtaining a new observation.

(3) Update the weights $W_{ij}$ by:

$$W_{ij}(t+1) = W_{ij}(t) - \alpha \frac{\partial E}{\partial W_{ij}} \tag{21}$$

where $\alpha$ is a learning rate (as in equations 8 and 11), and where now define:

$$E = \tfrac{1}{2}(e(t))^2, \tag{22a}$$

where:

$$e(t) = \hat{J}(\underline{R}(t+1))/(1+r) + U(t) - \hat{J}(\underline{R}(t)) - U_0 \tag{22b}$$

<u>This specification is incomplete</u>, because it does not specify how to compute the derivatives which appear in equation 21. This is not just a matter of computation; there is also a need to specify <u>which</u> partial derivatives or gradient we will use. In HDP proper, we calculate:

$$\frac{\partial E}{\partial W_{ij}} = 2e\left(\frac{\partial E}{\partial W_{ij}}\right) = 2e\left(-\frac{\partial}{\partial W_{ij}}\hat{J}(\underline{R}(t),W)\right) \tag{23}$$

This calculation treats $J(\underline{R}(t))$ in equation 22 as a function of W, but does not account for the effect of W on $J(\underline{R}(t+1))$! If one were solving equation 12 using the obvious extension of the Galerkin method [27], we would adapt the weights by use of the complete <u>total</u> gradient:

$$\frac{\partial E}{\partial W_{ij}} = 2e\left(-\frac{\partial}{\partial W_{ij}}\hat{J}(\underline{R}(t),W) + \frac{1}{1+r}\frac{\partial}{\partial W_{ij}}\hat{J}(\underline{R}(t+1),W)\right) \tag{24}$$

Combining equations 21, 22 and 23, pure HDP may be written as:

$$\Delta W = -2\alpha e(\nabla_p e) = 2\alpha e \nabla \hat{J}(\underline{R}(t)) \tag{25}$$

and the Galerkinized version of HDP, HDPG, may be written as:

$$\Delta W = -2\alpha e(\nabla_t e) = 2\alpha e(\nabla \hat{J}(\underline{R}(t)) - (1/(1+r))\nabla \hat{J}(\underline{R}(t+1))) \quad , \tag{26}$$

where $\nabla_p$ refers to the <u>partial</u> gradient with respect to the weights, and where $\nabla_t$ refers to the <u>total</u> gradient. All of these derivatives can be obtained very efficiently by use of backpropagation, using local parallel calculations and so on [8,26].

In DHP, we follow the <u>same three steps</u>, but we now define the error E as:

$$E = (\underline{e}(t))^2 = \sum_i (\lambda_i^*(t) - \hat{\lambda}_i(t))^2, \qquad (27)$$

where the <u>target value</u> $\lambda_i^*(t)$ is calculated by:

$$\lambda_i^*(t) = \frac{\partial U(t)}{\partial R_i} + \frac{1}{1+r}\sum_j \hat{\lambda}_j(t+1) \cdot \frac{\partial f_j(\underline{R}(t), \underline{w}(t))}{\partial R_i(t)} \qquad (28)$$

Note that $\hat{\lambda}_i(t)$ is simply the i-th component of the vector $\underline{\lambda}(\underline{R}(t),W)$ calculated by our Critic network, and that $\hat{\lambda}_j(t+1)$ is the j-th component of $\underline{\hat{\lambda}}(\underline{R}(t+1),W)$. Equation 28 corresponds to equation 10 of [8,ch.13] in the case where there is no Action network (and where the typo on the left is fixed); however [8,ch.13] explicitly indicates how the derivatives may be calculated efficiently, using "dual subroutines" or "dual functions" in order to implement backpropagation. Note as well that the backpropagation equations for equation 28 require that we <u>know the vector $\underline{w}(t)$</u>; when $\underline{R}(t+1)$ was obtained from simulation, we already know $\underline{w}(t)$, but when $\underline{R}(t+1)$ is observed directly, we can deduce $\underline{w}$ efficiently only if the Model network has a special kind of structure. (For example, the Model network could be based on the usual assumption that $R_i(t+1) = \hat{R}_i(t+1) + w_i$, or it could be based on a structure like the Stochastic Encoder/Decoder/Predictor of [8,ch.13].)

Here, as with HDP, we may choose between a pure version of DHP, based on $\nabla_p E$, or a Galerkinized version, based on $\nabla_t E$. In pure DHP:

$$\Delta W = -2\alpha \sum_k e_k \nabla_p e_k = 2\alpha \sum_k e_k \nabla \hat{\lambda}_k(\underline{R}(t)) = -2\alpha(\underline{e}^T \nabla_p \underline{e}) = 2\alpha(\underline{e}^T \nabla \underline{\hat{\lambda}}(t)) \qquad (29)$$

In DHPG:

$$\begin{aligned}
\Delta W_{ij} &= -2\alpha \sum_k e_k \left( \frac{\partial \lambda_k^*(t)}{\partial W_{ij}} - \frac{\partial \hat{\lambda}_k(t)}{\partial W_{ij}} \right) \\
&= -2\alpha \sum_k e_k \left( -\frac{\partial \hat{\lambda}_k(t)}{\partial W_{ij}} + \frac{1}{1+r} \sum_v \frac{\partial f_v(R(t), w(t))}{\partial R_k} \cdot \frac{\partial \hat{\lambda}_v(t+1)}{\partial W_{ij}} \right) \\
&= 2\alpha \sum_k e_k \frac{\partial \hat{\lambda}_k(t)}{\partial W_{ij}} - \frac{2\alpha}{1+r} \sum_{k,v} e_k \frac{\partial f_v}{\partial R_k} \cdot \frac{\partial \hat{\lambda}_v(t+1)}{\partial W_{ij}}
\end{aligned} \qquad (30)$$

This can be represented as:

$$\Delta W = -2\alpha \sum_k e_k \nabla_t e_k = 2\alpha(\mathbf{e}^T \nabla \hat{\underline{\lambda}}(t)) - \frac{2\alpha}{1+r} \mathbf{e}^T F^T(t) \nabla \hat{\underline{\lambda}}(t+1), \tag{31}$$

where F(t) is the Jacobian with respect to $\mathbf{R}(t)$ of the function $\mathbf{f}(\mathbf{R}(t),\mathbf{w}(t))$, and where all of the gradients are gradients with respect to the array of weights W.

Finally, in GDHP, we use the same three steps as with HDP in order to adapt the weights in the Critic network $J(\mathbf{R}(t),W)$, but we replace equation 22a with a much more complicated error function:

$$E = \Omega_0(e_0^2(t)) + \mathbf{e}^T \Omega \mathbf{e}(t), \tag{32}$$

where $\Omega_0$ is some nonnegative scalar and $\Omega$ is some positive definite matrix, where $e_0$ is defined to be the same as "e" as defined in equation 22b, and where $\mathbf{e}(t)$ is _essentially_ the same vector as in equation 27. In this case, of course, we cannot use exactly the same definition for $\mathbf{e}(t)$, because we do not have a $\hat{\underline{\lambda}}$ output available, when we are using a critic which outputs $\hat{J}$. Thus to define $\mathbf{e}(t)$ here, still using equations 27 and 28, we simply replace each reference to $\hat{\underline{\lambda}}$ by a reference to $\nabla \hat{J}$. This is already sufficient to define GDHP, in a purely mathematical sense. Again, of course, there are two versions of the method, one a pure version and the other a Galerkinized version, depending on whether one adapts the weights in proportion to $\nabla_p E$ or $\nabla_t E$.

GDHP is far more difficult to implement than HDP or DHP. Therefore, most researchers in this area focus their efforts on studying HDP and DHP, which already present major challenges and opportunities. Only Wunsch and Prokhorov have claimed a working implementation of GDHP[14]. To calculate the derivatives of E with respect to the weights, directly, we must use second-order backpropagation; the theory of how to do this is given in [8,ch.10], and pseudocode may be found in [71]. Indirect approaches may or may not converge to the same equilibrium; however, such issues are beyond the scope of this paper.
Even though GDHP is more difficult to implement than HDP or DHP, it is still a local, parallel neural network design, O(N) in cost in the same way as the other.

In [65], I proposed that $\Omega$ should be a diagonal matrix, in effect. From a neural network viewpoint, that is a reasonable restriction. For purposes of this paper, I have considered only two special cases: (1) $\Omega=I$; (2$\Omega$ as any positive symmetric real matrix.
In both cases, I have set $\Omega_0=0$. Strictly speaking, then, the conclusions below about GDHP apply only to these special cases of GDHP. But the general form of GDHP is a hybrid, in effect, of HDP (which minimizes $e_0^2$) and of those special cases; it is difficult to believe that any form of GDHP could be unconditionally stable if those special cases and HDP both fail to achieve unconditional stability.

### 4.3.3. Continuous-Time Versions of HDP, DHP and of GDHP in the Nonlinear/Neural Case

So far as I know, no one has ever formulated HDP, DHP or GDHP as methods for continuous-time adaptation. There are several reasons for this. First, in real-world computing, digital or analog, there always are time delays, especially when complex nonlinear calculations are required. Continuous-time equations which assume that special-purpose analog chips are used and that no such delays exist would not accurately represent the actual behavior of a real-world computational system. Second, there are fundamental theoretical reasons why higher-order intelligent systems <u>must</u> contain discrete "clocks" [32].
Third, there is evidence from a host of respected neuroscientists such as Llinas and Barry Richmond which indicates that there are surprisingly precise "clocks" all over the brain; for example, there are general, nonspecific synchronized signals which enter the cerebral cortex, and create alternating "windows" of 30-50 ms. in which the dynamics of the entire circuit appear to behave differently. Fourth, a discrete-time representation tends to simplify the communication and presentation of complex designs. Fifth, one can normally approximate the behavior of continuous-time systems with a series of discrete-time systems, but not vice-versa. Of course, the continuous time version of the Bellman equation -- sometimes called the Hamilton-Jacobi-Bellman equation -- has been well-known for decades [4].

On the other hand, Narendra [1] and Grossberg [72] and others have proved a number of very important stability theorems, using a combination of Liapunov stability analysis and continuous-time representations. There is an obvious fit between Liapunov theory and continuous-time representation. Because the purpose of this new work was to develop Liapunov stability results, intended to encourage later total system theorems in the spirit of Narendra, I began by developing a continuous-time representation of HDP, HDPC. I hoped that this representation would make it easier to perform rigorous stability analysis.
If stability had been proven that way, it would have been important to return later to consider the discrete-time case; however, stability of HDPC would have made a good starting point.

In actuality, I did not find really important differences between the discrete-time case and the continuous-time case for HDP. Intuitively, I now see little reason to expect such differences for DHP or GDHP, either. Nevertheless, this section will give continuous-time versions of all three methods, just for the record.

The continuous-time version of equation 15 is:

$$\partial_t J = - U + U_0 + rJ \qquad (33)$$

The continuous-time version of equation 21 is:

$$\partial_t W_{ij} = -\alpha \frac{\partial}{\partial W_{ij}} E = -2\alpha e \frac{\partial}{\partial W_{ij}} e, \qquad (34)$$

where:

$$e = \partial_t \hat{J} + U - U_0 - r\hat{J} \qquad (35)$$

For the continuous time version of pure HDP, HDPC, we use the partial gradient defined by:

$$\frac{\partial e}{\partial W_{ij}} = -\frac{\partial \hat{J}}{\partial W_{ij}} \qquad (36)$$

(Some would argue that we should multiply this by (1-r), but this would have no effect on equation 34 anyway, because it simply multiplies the arbitrary fixed scalar $\alpha$ by another fixed positive scalar.) For the continuous-time version of HDPG, HDPGC, we use the full gradient:

$$\frac{\partial e}{\partial W_{ij}} = \frac{d}{dt}\left(\frac{\partial \hat{J}}{\partial W_{ij}}\right) - r\frac{\partial \hat{J}}{\partial W_{ij}} \qquad (37)$$

where the total derivative with respect to time is computed as if W were constant; in other words, we define:

$$\frac{d}{dt}\left(\frac{\partial \hat{J}}{\partial W_{ij}}\right) = \sum_k \left(\frac{\partial^2 \hat{J}}{\partial W_{ij} \partial R_k} \dot{R}_k\right) \qquad (38)$$

which could be calculated in a number of ways (such as direct observation of the time-changes of the network under fixed weights or the use of second-order backpropagation).

The continuous-time version of DHP is based on the gradient with respect to **R** of equation 33:

$$\partial_t \underline{\lambda} = -\nabla U + r\underline{\lambda} \qquad (39)$$

The continuous-time version of equation 21, for E defined as in equation 27, is:

$$\partial_t W_{ij} = -\alpha \frac{\partial}{\partial W_{ij}}\left|\underline{e}^2\right| = -2\alpha \sum_k e_k \frac{\partial}{\partial W_{ij}} e_k \qquad (40)$$

where:

$$e_k = \partial_t \hat{\lambda}_k + \frac{\partial U}{\partial R_k} - r\hat{\lambda}_k \qquad (41)$$

For the continuous-time version of pure DHP, DHPC, we disambiguate equation 40 by defining:

$$\frac{\partial}{\partial W_{ij}} e_k = -\frac{\partial}{\partial W_{ij}} \hat{\lambda}_k \qquad (42)$$

For DHPGC, we instead define:

$$\frac{\partial}{\partial W_{ij}} e_k = \frac{d}{dt}\left(\frac{\partial \hat{\lambda}_k}{\partial W_{ij}}\right) - r\frac{\partial \hat{\lambda}_k}{\partial W_{ij}} \qquad (43)$$

where the total derivative with respect to time in equation 43 is defined exactly in parallel with the similar equations 37 and 38.

Finally, in GDHPC and GHDPGC, we adapt the weights according to:

$$\partial_t W_{ij} = -\alpha \frac{\partial}{\partial \mathbf{W}_{ij}} E ,  \qquad (44)$$

where E is defined as in equation 32, and where $e_0$, **e**(t) and their derivatives are based on e, **e**(t) and their derivatives for HDPC and DHPC -- exactly in parallel to the way that HDP and DHP are used in GDHP in the discrete-time case.

### 4.3.4. Previous Results on the Convergence of These Adaptation Rules

Issues of stability and convergence have been a major concern of the adaptive critic community for a long time. Nevertheless, formal results about MBAC systems have been relatively sparse. There have been the Ph.D. theses of Prokhorov [15] and of Landelius [43],
and unpublished results from Richard Saeks of Accurate Automation. The Saeks work focuses on whole-system stability issues, rather than Critic adaptation as such. Whole-systems stability is the major concern in [15] and [43], but they also consider Critic adaptation as such. Landelius mainly proves stability and convergence for his own Critic implementations, which were discussed in section 3. Prokhorov [15] proves a whole-systems stability result for an offline learning version of HDPG+BAC in the deterministic case; he also reports considerable empirical testing of many approaches, as will be discussed in section 8 and 9.

Tsitsiklis and Van Roy [73] have reviewed a large number of papers from the literature on lookup-table Critics, showing many specific examples where TD -- a special case of HDP -- becomes unstable; however, they prove that TD with lookup tables is stable, under certain restrictive assumptions. The key assumption, which they discuss at length, is that the training examples **x**(t) must be taken from the probability distribution implied by the Markhov chain being studied; stability against arbitrary training sequences is not guaranteed, even in the case of lookup table Critics. This paper will discuss a different class of plants to be controlled, but the conclusions about instability, at least, are consistent with those of Tsitsiklis
and Van Roy.

In addition to these important but sparse theoretical results, considerable practical energy has gone into the effort to make these methods converge, for a variety of different test problems, as mentioned in section 4.1. This has been a very important practical issue, even in the context of offline learning.

Some of this empirical work focused on the interplay between Action networks, Model networks and Critics. Narendra in [41] addresses related issues. But this paper focuses on Critic networks as such.

Among the tricks which are relevant to Critics as such, which have proven to be very useful in obtaining convergence, are:

(1) "<u>Shaping</u>" [7,8,453]. In shaping, one first adapts a network to solve one class of control problems, and then uses the resulting weights as the <u>initial values</u> of the weights of a network trained to solve similar but more difficult problems. When this method is applied over and over again, it provides a way of doing "step-by-step learning" (**not** the same as real-time or incremental learning!) analogous to the way that humans really learn to perform difficult tasks.

(2) <u>Interest rate management</u>. In this approach, one starts out with very large values of r, and then one gradually decreases r to zero. (Usually r=0 represents our true values, if the utility function is properly articulated[12,74].) Strictly speaking, this is just a special case of shaping, because different values of r represent different optimization problems. Large values of r represent shorter-term optimization problems, which are usually easier to solve than the long-term problems.

(3) <u>Utility function management</u>. This can also be viewed as a form of shaping, in a way. In some cases, the choice of utility functions has even been used as way of cheating, of injecting prior knowledge about the desired controller into the adaptive system. However, for reasons beyond the scope of this paper [10,11,674], it does make sense to think of a hybrid man-machine control system, in which the human (as the upper controller) passes on something like <u>his</u> learned J function to the lower controller (the machine), which treats those inputs as its fundamental values (U). As a practical example -- when we use gradient-based MBAC methods to search for optimal energy-saving "chaos controllers," it makes sense to use a relatively flat cost function U, representing energy consumption, in some large acceptable region of state space, and then to add a gradually rising penalty function (quadratic?) for points in some "buffer zone" between that region and regions which are truly dangerous.

These three tricks have been successful in a very wide range of studies, far too numerous for me to cite in detail. For example, one might look at the MBAC work cited in section 4.1. Also, many researchers in ADP for discrete state space reported similar results, in discussions organized by Sridhar Mahadevan of the University of South Florida, in connection with an NSF workshop in April 1996 on reinforcement learning oriented towards AI.

## 4.4. HDP, DHP and GDHP (Pure and Galerkinized) Applied to Linear Systems

The initial goal of this new work was to find out whether any of the methods discussed in section 4.3 could adapt a Critic in strongly stable, correct manner in the linear multivariate case. The first step in doing that work was to derive the details of those methods for that special case. This subsection will report the results of that first step.

### 4.4.1. Description of the Plants to be Controlled

As in section 4.3, we will assume that the Action network is held fixed, which results in an augmented model $\underline{R}(t+1) = \underline{f}(\underline{R}(t), \underline{w}(t))$. In working with the linear observable case, I will revert back to notation closer to that of Narendra and Annaswamy [1]; thus the state vector will be denoted simply as $\underline{x}$. Under these conditions, a linear deterministic MIMO plant obeys the following very simple dynamics:

$$\partial_t \underline{x}(t) = A\underline{x} \qquad (45)$$

for the continuous-time case, or:

$$\underline{x}(t+1) = A\underline{x}(t) \qquad (46)$$

for the discrete-time case. The linear stochastic MIMO plant obeys:

$$\partial_t \underline{x} = A\underline{x} + \underline{w} \quad , \tag{47}$$

in the continuous-time case, where $\underline{w}$ is a continuous stochastic process [4,5]. In the discrete-time case, it obeys:

$$\underline{x}(t+1) = A\underline{x}(t) + \underline{w}(t) \quad , \tag{48}$$

where $\underline{w}$ is a random vector based on the multivariate normal distribution $N(0,Q)$, which implies that:

$$<\underline{w}> = \underline{0} \tag{49}$$

$$<\underline{w}\underline{w}^T> = Q \tag{50}$$

Because our immediate goal (from section 3.2) is to minimize tracking error, which is a quadratic in the state vector, I assume here a quadratic cost function (technically, a quadratic disutility function) to be minimized:

$$U(\underline{x}) = \underline{x}^T V \underline{x} \tag{51}$$

Likewise I assume r=0.

### 4.4.2. General Characteristics of the Critics and Adaptation

For the linear-quadratic optimization problems of section 4.4.1, it is well known that the true J function of dynamic programming will be quadratic in the state vector $\underline{x}$. This was crucial to the research strategy discussed in section 3.2. Thus for all of the $\hat{J}$ Critic networks applied to these problems, I used:

$$\hat{J}(\underline{x}) = \underline{x}^T C \underline{x} \tag{52}$$

For the $\hat{\underline{\lambda}}$ Critics, I used the gradient of $\hat{J}$ with respect to $\underline{x}$:

$$\hat{\underline{\lambda}}(\underline{x}) = 2C\underline{x} \tag{53}$$

In order to analyze stability properties, I define a weight error matrix M:

$$M = C - C^* \tag{54}$$

where $C^*$ is the correct Critic weight matrix, the matrix which satisfies the Bellman equation for these systems:

$$J(\underline{x}(t)) = U(\underline{x}(t)) + <J(\underline{x}(t+1))> - U_0 \tag{55}$$

which, upon substitution from 51 and 53, reduces to:

$$C^* = V + A^T C^* A \tag{56}$$

both in the deterministic case (where $U_0=0$) and in the stochastic case
(where $U_0=\langle\underline{w}^T C *\underline{w}\rangle=Tr(C*Q)$, a scalar independent of $\underline{x}$).

For the stochastic version of these problems, with r=0, the $U_0$ term is not generally equal to zero. This affects HDP and its variants, but not DHP or GDHP (as implemented here). For simplicity, I did not consider the issue of adapting $U_0$ here, because it does not affect the fundamental issues under study here. In practical computing work where r=0, the usual practice is to start with r>0 and gradually send r towards zero, and see what the <u>Action network</u> converges to; the adaptation of $U_0$ has not received serious attention. This is a point worth returning to, as our basic understanding becomes more rigorous.

In describing the effects of adaptation in the discrete-time case, I have implicitly assumed infinitely small learning rates. Many techniques exist to reduce learning rates, in practice, when they are so large that they lead to instabilities. Those issues are certainly
worthy of study. Here, however, I have focused on the more basic question of whether these methods could be unconditionally stable, even if there were no danger of learning rates being set too high. This approach simplifies the analysis very greatly, because it lets us represent the adaptation of C as a <u>continuous</u> process represented as a differential equation. This, in turn, allows the use of the usual continuous-time Liapunov stability methods.

### 4.4.3. Details of HDPC and HDPGC

To derive the adaptation rule for C with HDPC and HDPGC in the linear case with r=0, we first substitute equations 51 and 52 into equation 35, which yields:

$$e = \partial_t(\underline{x}^T C \underline{x}) + \underline{x}^T V \underline{x} - U_0 = \underline{\dot{x}}^T C \underline{x} + \underline{x}^T C \underline{\dot{x}} + \underline{x}^T V \underline{x} - U_0 , \quad (57)$$

where "$\partial_t$" is a standard notation from physics for differentiation with respect to time.
For the deterministic case, we substitute equation 45 into equation 57, and recall from section 4.4.2 that $U_0=0$ here, which yields:

$$e = \underline{x}^T A^T C \underline{x} + \underline{x}^T C A \underline{x} + \underline{x}^T V \underline{x} \quad (58)$$

For HDPC, we substitute equation 52 into equation 36, remembering that the weights "$W_{ij}$" for the Critic shown in equation 52 are simply the matrix elements $C_{ij}$. This yields:

$$\frac{\partial e}{\partial C_{ij}} = -x_i x_j \quad (59)$$

Finally, substituting equations 59 and 58 into equation 34, we obtain the adaptation rule for HDPC in the linear deterministic case:

$$\partial_t C_{ij} = \alpha(\underline{x}^T A^T C \underline{x} + \underline{x}^T C A \underline{x} + \underline{x}^T V \underline{x}) x_i x_j \quad (60a)$$

which can be expressed more compactly as:

$$\partial_t C = \alpha e \underline{x}\underline{x}^T \quad (60b)$$

For HDPGC, it is easiest to compute the total gradient of e with respect to C by differentiating equation 58 directly:

$$V_t e = \underline{xx}^T A^T + A\underline{xx}^T \tag{61}$$

Substituting equations 61 and 58 back into equation 38, we get:

$$\partial_t C = -\alpha e(\underline{xx}^T A^T + A\underline{xx}^T) \tag{62}$$

In order to analyze the error dynamics here, we may first solve equation 54 to get:

$$C = M + C^* \tag{63}$$

Substituting this into equation 58, and subtracting out the terms containing C* (which cancel out due to the Bellman equation, as expressed in equations 55 and 56), we deduce:

$$e = \underline{x}^T A^T M \underline{x} + \underline{x}^T M A \underline{x} \tag{64}$$

Since the matrix C* in equation 63 is constant (i.e. unchanged by our adaptation process), substitution into equation 60 yields:

$$\partial_t M = \alpha e \underline{xx}^T \tag{65}$$

for HDPC, and:

$$\partial_t M = -\alpha e(\underline{xx}^T A^T + A\underline{xx}^T) \tag{66}$$

for HDPGC.

In this preliminary work, I did not analyze the behavior of HDPC or HDPGC in the linear stochastic case.

For all the varieties of HDP, we may assume that the matrices C*, C and M are symmetric, without loss of generality, because: (1) only the symmetric part of C or C* affects the function $J(\underline{x})$; (2) the update rules for C and M all involve the addition of symmetric matrices only (which can be seen by simply looking at the right-hand-side of these update equations).

### 4.4.4. Details of HDP and HDPG

For HDP and HDPG, I first analyzed the deterministic case, in parallel with section 4.4.3. Substituting equations 51 and 52 into 22b, again with r=0, we get:

$$e = \underline{x}(t+1)^T C \underline{x}(t+1) + \underline{x}^T V \underline{x} - \underline{x}^T C \underline{x} - U_0 \tag{67}$$

where I use the convention that "$\underline{x}$" refers to $\underline{x}(t)$. Substituting equation 46 into equation 67, and exploiting the fact that $U_0=0$ in the deterministic case (see section 4.4.2), we obtain:

$$e = \underline{x}^T A^T C A \underline{x} + \underline{x}^T V \underline{x} - \underline{x}^T C \underline{x} = \underline{x}^T (A^T C A + V - C) \underline{x} \tag{68}$$

For HDP, we substitute equation 52 into equation 23, to get:

$$\frac{\partial E}{\partial W_{ij}} = -2e x_i x_j \tag{69}$$

If we assume infinitely small learning rates in equation 21, such that equation 21 is converted into a differential equation (as discussed in section 4.4.2), the substituting from equation 69 into that equation yields:

$$\partial_t C_{ij} = -\tfrac{1}{2}\frac{\partial E}{\partial \mathbf{W}_{ij}} = ex_i x_j \qquad (70)$$

Note that I have selected an apparent learning rate of α=½; in actuality, the choice of apparent learning rate has no effect on the meaning of this equation, because it only rescales the arbitrary time scale of this equation. Because the apparent learning rate has no real effect, I simply choose a value which makes the equations appear as simple as possible.

For HDPG, we substitute equation 52 into equation 24, and then substitute the result into equation 21 in the same manner, which results in:

$$\partial_t C = -e(A\mathbf{xx}^T A^T - \mathbf{xx}^T) \qquad (71)$$

Substituting equation 63 into equation 68, and again exploiting the Bellman equation, we derive:
$$\mathbf{e} = \mathbf{x}^T(A^T M A - M)\mathbf{x} \qquad (72)$$

Substituting equation 63 into equation 70, we obtain:

$$\partial_t M = e\mathbf{xx}^T \qquad (73)$$

for HDP, and:

$$\partial_t M = e(\mathbf{xx}^T - A\mathbf{xx}^T A^T) \qquad (74)$$

for HDPG.

For the linear stochastic case, we substitute equation 48 into equation 57, and recall from section 4.4.2 that $U_0=\text{Tr}(CQ)$ here, which yields:

$$e = (A\mathbf{x} + \mathbf{w})^T C(A\mathbf{x} + \mathbf{w}) + \mathbf{x}^T V \mathbf{x} - \mathbf{x}^T C \mathbf{x} - \text{Tr}(C^*Q)$$

$$= \mathbf{x}^T A^T C A \mathbf{x} + \mathbf{x}^T A^T C \mathbf{w} + \mathbf{w}^T C A \mathbf{x} + \mathbf{w}^T C \mathbf{w} + \mathbf{x}^T V \mathbf{x} - \mathbf{x}^T C \mathbf{x} - \text{Tr}(C^*Q) \qquad (75)$$

For HDP, we again substitute equation 52 into equation 23, as in the deterministic case; the adaptation rule for C is still equation 70, but the meaning of equation 70 is changed by the use of a new definition for e, the definition given in equation 75. However, if we take statistical expectations, with respect to the random vector $\mathbf{w}$, we get:

$$\langle \partial_t C \rangle = \langle e\mathbf{xx}^T \rangle = \mathbf{xx}^T \langle e \rangle$$

$$= \mathbf{xx}^T (\mathbf{x}^T A^T C A \mathbf{x} + \mathbf{x}^T V \mathbf{x} - \mathbf{x}^T C \mathbf{x} + \langle \mathbf{x}^T A^T C \mathbf{w} \rangle + \langle \mathbf{w}^T C A \mathbf{x} \rangle + \langle \mathbf{w}^T C \mathbf{w} - \text{Tr}(C^*Q) \rangle)$$

$$= \mathbf{xx}^T (\mathbf{x}^T A^T C A \mathbf{x} + \mathbf{x}^T V \mathbf{x} - \mathbf{x}^T C \mathbf{x} + \text{Tr}((C - C^*)Q)) \qquad (76)$$

In deriving equation 76, I have exploited equations 49 and 50 in effect.

For HDPG in the stochastic case, we first differentiate equation 75 with respect to C:

$$\nabla_t e = A\underline{xx}^T A^T + \underline{wx}^T A^T + A\underline{xw}^T + \underline{ww}^T - \underline{xx}^T \tag{77}$$

Substituting into equation 24 and then 21, we arrive at:

$$\partial_t C = -e(A\underline{xx}^T A^T + \underline{wx}^T A^T + A\underline{xw}^T + \underline{ww}^T - \underline{xx}^T) \tag{78}$$

The expectation value of equation 78 is somewhat nasty to calculate, because it involves products up to the fourth power in the random vector $\underline{w}$. However, in the definition of e in equation 75, we can see that four of the terms do not depend upon $\underline{w}$; pulling out their contribution, and then exploiting the fact that the expectation value of odd powers of $\underline{w}$ will always equal zero, we can expand equation 78 to:

$$\langle \partial_t C \rangle = -(\underline{x}^T(A^T CA + V - C)\underline{x} - \text{Tr}(C*Q))(\langle A\underline{xx}^T A^T + \underline{wx}^T A^T + A\underline{xw}^T + \underline{ww}^T - \underline{xx}^T \rangle)$$

$$- \langle (\underline{x}^T A^T C\underline{w} + \underline{w}^T CA\underline{x})(\underline{wx}^T A^T + A\underline{xw}^T) \rangle$$

$$- \langle (\underline{w}^T C\underline{w})(A\underline{xx}^T A^T + \underline{ww}^T - \underline{xx}^T) \rangle \tag{79}$$

The middle line of this equation is somewhat tricky to expand. As an example, one of its terms, $\langle (\underline{x}^T A^T C\underline{w})\underline{wx}^T A^T \rangle$, can be expressed in an Einstein-like convention, in which doubled subscripts are implicitly summed over, yielding:

$$\langle (x_i A^T_{ij} C_{jk} w_k) w_\mu x_\lambda A^T_{\lambda\nu} \rangle = x_i A^T_{ij} C_{jk} Q_{k\mu} x_\lambda A^T_{\lambda\nu} \tag{80}$$

which represents the matrix:

$$QCA\underline{xx}^T A^T \tag{81}$$

Using this kind of approach, we may expand equation 79 further:

$$\langle \partial_t C \rangle = -(\underline{x}^T(A^T CA + V - C)\underline{x} - \text{Tr}(C*Q))(A\underline{xx}^T A^T + Q - \underline{xx}^T)$$

$$- 2QCA\underline{xx}^T A^T - 2A\underline{xx}^T A^T CQ - \text{Tr}(CQ)(A\underline{xx}^T A^T - \underline{xx}^T) - \langle (\underline{w}^T C\underline{w})(\underline{ww}^T) \rangle$$

$$= -(\underline{x}^T(A^T CA + V - C)\underline{x} + \text{Tr}((C-C*)Q))(A\underline{xx}^T A^T + Q - \underline{xx}^T)$$

$$- 2QCA\underline{xx}^T A^T - 2A\underline{xx}^T A^T CQ + \text{Tr}(CQ)Q - \langle (\underline{w}^T C\underline{w})(\underline{ww}^T) \rangle \tag{82}$$

Finally, for the analysis of error dynamics, we may again substitute equation 63 into equations 76 and 82, and exploit the Bellman equation. For HDP, we get:

$$\langle \partial_t M \rangle = \underline{xx}^T (\underline{x}^T(A^T MA - M) + \text{Tr}(MQ)) \tag{83}$$

For HDPG, we get:

$$\langle \partial_t M \rangle = -(\underline{x}^T(A^T MA - M)\underline{x} + \text{Tr}(MQ))(A\underline{xx}^T A^T + Q - \underline{xx}^T) \tag{84}$$

$$- 2Q(M+C*)A\underline{xx}^T A^T - 2A\underline{xx}^T A^T(M+C*)Q$$

$$+ \text{Tr}((M+C^*)Q)Q - \langle(\underline{w}^T(M+C^*)\underline{w})(\underline{w}\underline{w}^T)\rangle$$

Equation 84 could be expanded further, but it contains enough information for the analysis required later on in this paper.

Again, as in section 4.4.3, we can assume without loss of generality that C, C* and M are all symmetric matrices.

### 4.4.5. Details of DHP and DHPG

In order to derive the details of DHP and DHPG in the stochastic linear case, with r=0, we first substitute from equations 51 and 48 into equation 28, to deduce:

$$\underline{\lambda}^* = 2V\underline{x} + A^T\hat{\underline{\lambda}}(t+1) \tag{85}$$

Substituting 53 and 48 into 85, we get:

$$\underline{\lambda}^* = 2V\underline{x} + A^T(2C)((A\underline{x}+\underline{w}) = 2V\underline{x} + 2A^TCA\underline{x} + 2A^TC\underline{w} \tag{86}$$

Substituting 53 and 86 into 27, we deduce:

$$\underline{e} = \underline{\lambda}^* - \hat{\underline{\lambda}} = 2(V\underline{x} + A^TCA\underline{x} + A^TC\underline{w} - C\underline{x}) \tag{87}$$

To derive the details for DHP in this application, we next calculate the derivatives which make up the partial gradient, simply by differentiating equation 87 directly:

$$\frac{\partial}{\partial C_{ij}} e_k = 2(-\delta_{ik}x_j) \; ! \tag{88}$$

From this, one may easily deduce that:

$$(\underline{e}^T \frac{\partial}{\partial W_{ij}} \underline{e}) = \sum_k e_k \frac{\partial}{\partial C_{ij}} e_k = -2e_i x_j \tag{89}$$

Substituting equations 87 and 89 into equation 29, and converting equation 29 into a differential equation (in parallel with what we did in section 4.4.4), we obtain:

$$\partial_t C = -\tfrac{1}{2}(\underline{e}^T \nabla_p \underline{e}) = (V\underline{x} + A^TCA\underline{x} + A^TC\underline{w} - C\underline{x})\underline{x}^T$$

$$= (V + A^TCA - C)\underline{x}\underline{x}^T + A^TC\underline{w}\underline{x}^T \tag{90}$$

The expected value over the random vector $\underline{w}$ is easily calculated:

$$\langle\partial_t C\rangle = (V + A^TCA - C)\underline{x}\underline{x}^T \tag{91}$$

Note that the deterministic linear case is simply the special case where **w**=**0**;
in that case, $\partial_t C = <\partial_t C>$ is still given by equation 91.

In order to derive the details for DHPG, we first determine the values of two of the terms which appear in equation 30, for the linear stochastic case:

$$\frac{\partial f_v}{\partial R_k} = A_{vk} \tag{92}$$

$$\frac{\partial \hat{\lambda}_v(t+1)}{\partial W_{ij}} = \frac{\partial}{\partial C_{ij}}(2C(A\underline{x}+\underline{w}))_v = 2\delta_{iv}(A\underline{x}+\underline{w})_j \tag{93}$$

We may then substitute into the more complicated summation in equation 30:

$$\sum_{k,v} e_k \frac{\partial f_v}{\partial R_k} \cdot \frac{\partial \hat{\lambda}_v(t+1)}{\partial W_{ij}} = 2\sum_{k,v} e_k A_{vk} \delta_{iv}(A\underline{x}+\underline{w})_j = 2(A\underline{e})_i (A\underline{x}+\underline{w})_j \tag{94}$$

Substituting back into equation 30, and translating these update equations into a differential equation as before, we get:

$$\partial_t C = \underline{e}\underline{x}^T - A\underline{e}(\underline{x}^T A^T + \underline{w}^T) \tag{95}$$

$$= (V + A^T CA - C)\underline{x}\underline{x}^T + A^T C\underline{w}\underline{x}^T - A(V + A^T CA - C)\underline{x}\underline{x}^T A^T$$

$$- AA^T C\underline{w}\underline{x}^T A^T - A(V + A^T CA - C)\underline{x}\underline{w}^T - AA^T C\underline{w}\underline{w}^T$$

This implies:

$$<\partial_t C> = (V + A^T CA - C)\underline{x}\underline{x}^T - A(V + A^T CA - C)\underline{x}\underline{x}^T A^T - AA^T CQ \tag{96}$$

In order to analyze the error dynamics of pure DHP, we substitute equation 63 into equation 91, and again use the Bellman equation, to derive:

$$<\partial_t M> = (A^T MA - M)\underline{x}\underline{x}^T \tag{97}$$

For DHPG, the same kind of substitution into equation 96 yields:

$$<\partial_t M> = (A^T MA - M)\underline{x}\underline{x}^T - A(A^T MA - M)\underline{x}\underline{x}^T A^T - AA^T(C+M)Q \tag{98}$$

Note that the update rules shown in equations 91 and 96-98 <u>do not in general add a symmetric matrix</u> to C or M! With pure DHP or with DHPG, we cannot safely assume that C or M is a symmetric matrix; even if OUR <u>initial</u> estimate of C* is symmetric, the updated estimates after learning will usually not be symmetric. This is a valid consequence of the original nonlinear Critic adaptation rule which these special cases are being derived from. It is a manifestation of a general problem in the nonlinear case: the problem that DHP does not force **λ**(**R**(t)) to be an integrable vector field, even though we know that the gradient of any scalar field (like J($\underline{\mathbf{R}}$(t)) should be integrable. This problem was one of the main motivations behind the development of GDHP, which guarantees integrability even in the nonlinear case.

### 4.4.6. Details of GDHP and GDHPG

GDHP is more complicated than HDP or DHP in the general nonlinear case, as discussed in section 4.3.2. However, in the linear stochastic case, with $\Omega_0=0$ (which allows us to ignore $U_0$), it reduces to something relatively simple.

Let us begin by imagining that C might or might not be symmetric, for the sake of generality. For the case where $\Omega_0=0$ and $\Omega=I$, equation 32 becomes exactly the same as equation 27, except that we now use alternative sources for the vectors $\underline{\lambda}^*$ and $\hat{\underline{\lambda}}$. As discussed in section 4.3.2, we define these vectors by using $\nabla J$ in place of equation 53. Differentiating J as given in equation 52, we get:

$$\underline{\lambda}(\underline{\mathbf{x}}) = (C + C^T)\underline{\mathbf{x}} \quad (99)$$

Aside from this replacement, the further calculations for GDHP are identical to those for DHP in section 4.4.5. When we calculate derivatives of E or e with respect to the weights $C_{ij}$, equation 99 simply replaces the previous (DHP) value of $(\partial E/\partial C_{ij})$ by the average ½$((\partial E/\partial C_{ij})+(\partial E/\partial C_{ji}))$, because each weight $C_{ij}$ appears both in C (as before) and in $C^T$. The net effect of this is very simple: <u>all</u> of the adaptation equations for C, and dynamic equations for M, are exactly the same as in section 4.4.5, except that we divide every update by two, and add it to its transpose. (Also, the division by two can be skipped, since it simply multiplies everything by a scalar factor which can be absorbed into the learning rate.)

Thus for GDHP in the stochastic case, equation 91 is replaced by:

$$<\partial_t C> = (V + A^T CA - C)\underline{\mathbf{x}}\underline{\mathbf{x}}^T + \underline{\mathbf{x}}\underline{\mathbf{x}}^T(V + A^T CA - C) \quad (100)$$

Its error dynamics are:

$$<\partial_t M> = (A^T MA - M)\underline{\mathbf{x}}\underline{\mathbf{x}}^T + \underline{\mathbf{x}}\underline{\mathbf{x}}^T(A^T MA - M) \quad (101)$$

For GDHPG, the error dynamics follow the symmetrization of equation 98:

$$<\partial_t M> = (A^T MA - M)\underline{\mathbf{x}}\underline{\mathbf{x}}^T + \underline{\mathbf{x}}\underline{\mathbf{x}}^T(A^T MA - M) - A(A^T MA - M)\underline{\mathbf{x}}\underline{\mathbf{x}}^T A^T$$

$$- AA^T(C^*+M)Q - A\underline{\mathbf{x}}\underline{\mathbf{x}}^T(A^T MA - M)A^T - Q(C^*+M)AA^T \quad (102)$$

Strictly speaking, the asymmetry problem with DHP could have been patched up by using a neural network trick called "weight sharing" or "symmetry exploitation" -- which works here <u>specifically in the linear/quadratic case</u>. In my initial investigations, I did use that approach, which I called "DHPS" (DHP Symmetrized). But DHPS is identical to GDHP in the linear/quadratic case. GDHP, however, is a <u>general</u> nonlinear adaptation method.

I have also briefly considered the case where $\Omega$ may be a general positive symmetric matrix -- not necessarily I -- both as a generalization of DHP and for the symmetrized, GDHP version. For DHP, we simply replace equation 89 with:
AQ

$$(\underline{e}^T \Omega \frac{\partial}{\partial \mathbf{W}_{ij}} \underline{e}) = \sum_k (\Omega \underline{e})_k \frac{\partial}{\partial C_{ij}} e_k \qquad (103)$$

This simply tells us to replace the vector <u>e</u> on the left side of equation 90 with Ω<u>e</u>, which then changes equation 91 to:

$$<\partial_t C> = \Omega(V + A^T CA - C) \underline{xx}^T \qquad (104)$$

The resulting error dynamics are of course:

$$<\partial_t M> = \Omega(A^T MA - M)\underline{xx}^T \qquad (105)$$

For GDHP, the error dynamics are just the symmetrized version of equation 105:

$$<\partial_t M> = \Omega(A^T MA - M)\underline{xx}^T + \underline{xx}^T (A^T MA - M)\Omega \qquad (106)$$

## 5. A Pathway to Stable Universal Adaptive Control and Nonlinear Control

The goal of this paper is to lay the groundwork for a number of possible research directions. I hope that other researchers will reap the benefits of these opportunities, because they are far too large for any one individual to fulfill. This section will discuss three of the possible new directions: (1) universal adaptive control for the linear/quadratic case; (2) extensions for multiple models or piecewise linear, gain-scheduled control; (3) true intelligent control.

### 5.1. Linear/Quadratic Universal Adaptive Control

This opportunity -- the closest at hand -- was already discussed in section 3.2. The first opportunity or task is to design a universal adaptive controller for the MIMO linear deterministic case, guaranteed to stabilize (asymptotically) any unknown plant which is stabilizable; in other words, this universal adaptive controller should be able to stabilize any plant for which a fixed controller could stabilize (asymptotically) the plant if the dynamics of the plant were known.
    Such an achievement would require <u>total system stability theorems</u> in the spirit of Narendra and Annaswamy [1] and others. But before we attempt to attain or prove total system stability, we should first make sure that we have <u>individual components</u> -- Action, Model and Critic -- which are at least stable under the most benign conditions, the conditions where the other components are held fixed. Stable system identifiers and Action/controller systems already exist [1]; however, no one has ever provided Critic adaptation methods which meet the strict standards for stability which I will discuss in section 7.
    In summary, from a theoretician's point of view, this paper attempts to lay the groundwork which makes it possible to do total system stability analysis, with some chance of success, for the linear deterministic fully-observed case. This may sound like a very small start; however a true universal adaptive controller even for that case would be a major achievement. Following the approach in [1] (and perhaps drawing on [6]), an extension to the partially observed case would be one logical follow-on. Clearly the linear stochastic case is a reasonable goal as well, after the deterministic case is nailed down.
    In actuality, even this paper is asking for more than just a universal adaptive controller for the linear deterministic case; it is asking that that controller be based on a Critic adaptation

method which converges to the right answer in the stochastic case, derived from an adaptation rule which makes sense in the general nonlinear stochastic situation. Adhering to this constraint should make it easier to carry through the extensions to the stochastic and nonlinear cases in the more distant future.

From the practitioner's point of view, the new Critic adaptation rules to be discussed here are already useable as an upgrade to the existing methods, which are already relevant to a number of applications as discussed in [7] and in section 4.1.

**5.2. Multiple Models, Fuzzy IF-THEN Rules and Piecewise Linear Control**

Narendra has asserted in several contexts that multiple-model methods are a key to intelligent control. There are several lines of argument which support that viewpoint [13]. Nevertheless, I would argue that the complexity and the power of multiple models should not be wasted on patching up more elementary problems, like the minimum phase and sign-information problems in adaptive control. When a plant is known to be linear, and when the state space is not divided up into different regions governed by different laws, then single-model methods should suffice.

There is a more positive way to express this point: based on the discussion of section 3.2, there is good reason to believe that the present restrictions on adaptive control (minimum phase and knowledge of gain structure) can be overcome in the linear case, without resorting to multiple models. These problems -- as well as slow transient response -- are essentially the result of minimizing a simple measure of tracking error at time t+1. An approach which converges to the dynamic programming solution should converge to a stable and optimal solution which, like model-predictive-control, should also show good response to transient disturbances, when sufficient input data is available.

On the other hand, there are three possible uses of multiple models which remain valid and important:

(1) The use of multiple models to explicitly represent qualitative uncertainty about the world, when that uncertainty cannot be adequately represented by probability distributions over continuous [8,ch.13] or discrete [75] variables. This kind of intellectual strategy is very important to sophisticated symbolic reasoning in humans.

(2) The use of multiple models to characterize very different regions of state space [13]. Piecewise linear models of nonlinear systems are one classic example of this approach. Piecewise linear models or approximations can emerge in many different ways. For example, Sugeno's classic fuzzy controller for a helicopter (presented at the World Automation Congress 1998) is often cited as the most impressive success story for fuzzy control; it is based on fuzzy IF-THEN rules which translate down, in practice, to a piecewise linear system. Classical gain-scheduling embodies the same underlying principles, with less user-friendly interface. Multiple nonlinear models have also proven very useful and powerful, and are now very popular in parts of the neural network community [76].

(3) The use of multiple models to characterize different tasks or action schemata or decision blocks [9,13,46] in a hierarchical decision-making system.

Of these three directions, the application to piecewise linear control seems most relevant to the present state of the art in control engineering. Improved control within each linear region would clearly be very important to these kinds of schemes. The work proposed in section 5.1 should develop better control possibilities within each linear region; this, in turn, should make it easier to build more powerful control systems by linking together these multiple regions.

The existing literature on stable control for piecewise linear systems (as in [77] or in Sugeno's stability proofs) emphasizes the development of a <u>common</u> quadratic Liapunov function, applicable to <u>all</u> of the regions. When this works, it certainly yields stable controllers. However, it is obvious that there are many piecewise linear systems which can be stabilized in principle, in which one cannot use the same Liapunov function across all regions (For example, one region may be nonminimum phase, relative to the metric or coordinate system implied by the Liapunov function used in another region.) But how could we <u>construct</u> or <u>determine the existence</u> of piecewise quadratic Liapunov functions? (Or of whatever class of Liapunov function is guaranteed to work in piecewise linear systems which can be stabilized.) The first step in developing such constructive methods is to fully develop constructive methods for <u>individual</u> linear/quadratic regions; the research proposed in section 5.1 should help.

Going beyond that starting point, there may be significant challenges in weaving together different Critics for different regions, Critics which must interface properly at the boundaries between those regions. The new Werbos-Bellman equations given in [13] (extracted from a longer, denser paper [46]) provide a rigorous basis for managing that interface. The development of those equations was initially motivated by an effort to develop more truly brain-like intelligent control; however, the mathematics is not limited to that application.

### 5.3. The Search for True Intelligent Control

The research proposed above can be seen as one strand of a larger research effort, an effort which has the potential to replicate the kind of higher-level learning and intelligence which we see in the brains of mammals. In [9,46], I have argued that we now have the necessary conceptual tools in place to pursue that goal in a truly serious way, for the first time, with a great likelihood of success. There are at least six areas, however, where more research is needed:

(1) Even in the "simple" testbed problem of supervised learning (learning a static mapping from an input vector **X** to a vector of targets or dependent variables **Y**), we need improved learning speed and generalization ability, exploiting concepts such as "syncretism" (section 2.2) and "simultaneous recurrence" [32]. Also see [34].

(2) In adapting Model networks, we need to explore architectures (like SEDP [8,ch.13] and like [75]) which can represent <u>probability distributions</u>, not just deterministic input-output relations. We need to substantially improve "robustness over time" [8,ch.10]. We need to explore Error Critics [8, ch.13;26;15] as an alternative to BTT, in order to combine real-time learning with the robust kind of memory now available only with BTT. The improvements needed in supervised learning should be extended to Model training as well.

(3) Substantial improvements in basic ADP technology are also needed. The work proposed in this paper addresses a portion of those needs.

(4) At a higher level, we need to develop, understand, test and apply designs for systems which perform "temporal chunking" in an adaptive way [9,13,46].

(5) Better systems are needed for "brain-like stochastic search, in order to help decision-making system escape from local minima, without the artificial limitations of methods like genetic algorithms which, in my view, are not truly brain-like. For example, suppose that we face a class of static function maximization problems, in which each specific problem is characterized by set of parameters **α**; we are asked to find actions or decisions **u** which maximize $U(\boldsymbol{\alpha},\mathbf{u})$. (As an example, **α** might represent the locations of cities in a traveling salesman problem.) We can train an Option Network [46] so that it inputs **α** and outputs possible **u**, in a stochastic way, and tries to ensure that it outputs **u** according to a Gibbs distribution, c*exp(-(k/T)U), where T is a temperature parameter also input to the network.

(6) Principles like spatial symmetry and objective symmetry [46] need to be implemented in new classes of neural networks, in order to make it possible for these networks to cope with the huge volumes of input which biological organisms can handle. In effect, we need to work on neural adaptive designs which learn to do what some AI systems already do in a crude, brute-force manner: build up spatial "world models;" analyze objects one at a time, in a multiplexing kind of mode; perform spatial chunking; input and output networks of relations rather than fixed-length vectors. Again, see [46].

The research proposed in section 5.2 would contribute to area four as well as area three. These areas all interconnect, of course; optimal practical designs would normally bring together elements from many of these strands of research.
In general, I would agree with those critics who argue that the human mind has greater capabilities than the capabilities which this kind of research can provide [12]. The human mind is more than just a very big mouse brain. Even quantum effects may ultimately prove to be of serious value at the systems design level. Nevertheless, the effort to really understand the level of intelligence found in basic mammalian brains, in a mathematically sophisticated way, is already a difficult enough challenge for now. It may be a prerequisite to a deeper, less fuzzy understanding of the higher levels.

# 6. Which Methods Converge to The Right Answer?

## 6.1. Introduction and Summary

At the start of this work, I hoped that at least one of the methods specified in section 4.4 would meet both of the following basic tests:

(1) **"Correct equilibrium"**, defined as follows: when C equals the correct C* matrix, defined by equation 56, then I would ask that the adaptation methods results in $<\partial_t C> = 0$ in general in the linear stochastic case.

(2) Stable convergence to the equilibrium: I have hoped that the adaptation rule would possess a property which I call quadratic unconditional stability -- to be defined in section 7 -- in the linear deterministic case, at least.

Unfortunately, none of these methods passed both of these tests. In particular, none of the Galerkinized methods from section 4.4 possess the property of correct equilibrium. HDPC and HDPGC were not studied here, but I see no reason to expect that they would have been any better than HDP and HDPG; after all, they are imply the limiting case of HDP and HDPG
as the time interval goes to zero.
These results are consistent with earlier work [17], addressing the case of linear/linear multivariate plants; in that work HDP demonstrated correct equilibrium for stochastic plants, but HDPG did not.
This section will analyze the equilibrium properties of the discrete-time methods given in section 4.4. Section 7 will evaluate their stability properties in the linear deterministic case.,

Section 8 will discuss the fundamental issues involved, including the Generalized Moving Target (GMT) problem and new designs for that problem. Section 9 will propose new nonlinear designs whose linear/quadratic special case passes both tests. Again, these tests are intended only as a simple starting point for a larger strand of research.

**6.2. Equilibrium Properties of HDP, DHP and GDHP**

From the definition of M in equation 63, the property of "correct equilibrium" as defined above is equivalent to the property that $<\partial_t M> = 0$ when M=0.

For HDP, the dynamics of M in the stochastic case are given in equation 83. When we substitute M=0 into equation 83, it is obvious that the right-hand side will equal zero. Thus HDP possesses correct equilibrium, under the assumptions of section 4.4.

Likewise, the dynamics of M in the stochastic case for DHP are given in equation 97. Again, inserting M=0 sends the right-hand side to zero, which demonstrates correct equilibrium.

Finally, for GDHP, in the most general version analyzed in section 4.4, the dynamics of M in the stationary case are given in equation 105. Once again, if we insert M=0, we find that $<\partial_t M>=0$, as required. Thus GDHP also possesses correct equilibrium.

**6.3. Equilibrium Properties of HDPG, DHPG and GDHPG**

The equilibrium properties of these methods can again be analyzed in the same way that we did in section 6.2.

The dynamics of M for HDPG in the stochastic case are given in equation 84. When we substitute M=0 into that equation, we obtain:

$$<\partial_t M> = -2QC^*A\underline{xx}^T - 2A\underline{xx}^T A^T C^*Q + Tr(C^*Q)Q - <(\underline{w}^T C^* \underline{w})\underline{ww}^T> \quad (107)$$

If Q=0 (the deterministic case), this will still go to zero. This is like the results in [17], where it was shown that HDPG yields an incorrect equilibrium <u>almost always.</u> There was a small list of special cases, of plants where the equilibrium would be correct. At the top of that list was the case where Q=0. But beyond that, for normal plants we would not expect C* to be a singular matrix. In a stochastic system, we would not expect the state vectors $\underline{x}$ to always fit the condition $A\underline{x}=\underline{0}$ or (even less likely) to always have the property that they exactly cancel out the last two terms in equation 107. One could spend considerable time trying to spell out the exact rare circumstances (other than Q=0) which would allow equation 107 to be zero; however, it is clear that equation 107 will usually be different from zero, almost always, for a normal stochastic system. Certainly equation 107 is not <u>always</u> zero, as the property of "correct equilibrium" requires.

Likewise, the dynamics of M for DHPG in the stochastic case are given in equation 98. When we substitute M=0 into that equation, we obtain:

$$<\partial_t M> = -AA^T C^*Q \quad (108)$$

Once again, when Q is nonzero, we normally expect this matrix to be nonzero as well. Note that this equation indicates a constant, unremitting bias away from the correct equilibrium, regardless of the states $\underline{x}$ actually experienced.

For GDHPG (with $\Omega$=I), the error dynamics for the stochastic case are given in equation 102. Substituting M=0 yields:

$$<\partial_t M> = -AA^T C^*Q - QC^*AA^T \quad (109)$$

The conclusions here are exactly the same as for DHP.

In summary, none of the Galerkinized methods discussed in section 4.4 possess "correct equilibrium" in the linear stochastic case.

## 7. Which Methods Possess Quadratic Unconditional Stability?

### 7.1. Introduction and Summary

This section will attempt to answer the following question: which of the methods given in section 4.4 possess <u>quadratic unconditional stability</u> in the deterministic case?

From the start, the reader should be warned that this concept of stability is extremely stringent -- far more stringent than we need in practice. I have chosen to investigate it for two main reasons: (1) it is a relatively easy point of departure for an initial
phase of research; (2) when methods can be found which <u>do</u> meet this stringent test, they would tend to be of practical interest as well.

There are many, many concepts of stability useful in different types of application [1,73, 78]. The remainder of this paper will use the term "q-stability" for this particular concept. Section 8 will describe how adaptive methods can be very stable, in a practical sense, even without possessing this very special property. Systems which do not possess q-stability will be called "q-unstable."

Q-stability is defined as follows, for the kinds of adaptive systems discussed in section 4.4. For any dynamic process $\{A,Q,V\}$, an adaptation process for C is q-stable iff there exists a Liapunov function ) -- or, equivalently, $\Lambda(M,A,Q,V)$ -- such that:

(1) $\Lambda$ is a positive definite quadratic function of M;

(2) $<\partial_t \Lambda> \leq 0$ for all combinations of M and $\underline{x}$. (110)

In equation 110, the angle brackets refer to the expectation value with respect to the random vector $\underline{w}$.

Intuitively, unconditional stability requires stability in the face of <u>any</u> sequence of inputs $\underline{x}$ -- not just the sequences which the plant would normally generate if it obeys equations 45 or 46 (or even 47 or 48). In effect, we are asking the system to learn and thrive even when the experience it grows up on is "a movie made by a devil" -- a sequence of collages (pairs of $\underline{x}(t)$ and $\underline{x}(t+1)$), each of which flows naturally internally, but which fade out and fade in in an illogical manner, specifically chosen so as to destabilize the learning process. Q-stability is the special case of unconditional stability, in which the Liapunov function is quadratic; section 8 will give an example of a linear dynamic system which is unconditionally stable but not q-stable.

The notion of q-stability given above is almost identical to the notion of quadratic stability as discussed in the seminal book by Boyd et al [79]. On page 73 of that book, they note that it is possible for a Linear Differential Inclusion (like the dynamic rules for M given
in section 4.4) to be stable, without being quadratically stable. On page 74, they cite a number of examples; they state that unconditional stability in a more general sense occurs "... if and only if there is a convex Liapunov function that proves it... IN general, computing such Liapunov functions is computationally intensive, if not intractable."

Ideally, we would hope to find adaptive methods which are q-stable for all dynamic processes such that the eigenvalues of A lie within the unit circle. The Galerkinized methods do possess that property, for deterministic linear plants. But the other methods -- the methods which possess "correct equilibrium" (as discussed in section 6) -- are q-unstable, even in the deterministic case, except for a "set of measure zero" in the universe of dynamical systems. For practical purposes, they are q-stable only when A is a perfectly <u>symmetric</u> matrix.

As in section 6.2, I will not attempt to specify very precisely what the special conditions are which would allow these methods to work. It is enough to know that they fail the test almost always. There is room for more formal mathematicians to tighten up the arguments here; however, the basic conclusions seem inescapable.

## 7.2. A Warmup Example: LMS Learning

This section will illustrate the idea of q-stability by showing how it applies to a well-known, fundamental learning rule used in supervised learning and signal processing. It will address the real-time learning rule called Least Mean Squares (LMS) developed by Widrow and Hoff [80], in a linear multivariate case. The robust stability of LMS is already well-known [42]; my goal here is simply to prepare the reader for later subsections.

Let us assume that we are observing a sequence of pairs, **x**(t) and y(t), of a vector **x** and a scalar y governed by the stochastic process:

$$y(t) = (\underline{b}^*)^T \underline{x}(t) + w(t), \qquad (111)$$

where w is a normally distributed random variable. Let us assume a learning process based on:

$$\Delta b_i = -\alpha \frac{\partial E}{\partial b_i}, \qquad (112)$$

where we define:

$$E = \tfrac{1}{2} e^2 = \tfrac{1}{2}(y - \underline{b}^T \underline{x})^2 \qquad (113)$$

Substituting 113 into 112, we first get:

$$\frac{\partial E}{\partial b_i} = - (y - \mathbf{b}^T\mathbf{x})x_i \tag{114}$$

Assuming infinitely small learning rates (as discussed in section 4.4.2), we substitute equation 114 into equation 112, and represent the result as a differential equation:

$$\partial_t \mathbf{b} = (y - \mathbf{b}^T\mathbf{x})\mathbf{x} \tag{115}$$

Substituting equation 111 into this, we get:

$$\partial_t \mathbf{b} = (\mathbf{b}^T\mathbf{x} + w - \mathbf{b}^T\mathbf{x})\mathbf{x} \tag{116}$$

In parallel with equation 63, we may implicitly define a new vector $\boldsymbol{\delta}$:, which represents the error of our adaptive system in estimating the parameter vector **b:**

$$\mathbf{b} = \mathbf{b}^* + \boldsymbol{\delta} \tag{117}$$

Substituting this definition into equation 116, and taking expectation values with respect to the random variable w, we get:

$$<\partial_t \boldsymbol{\delta}> = - (\boldsymbol{\delta}^T\mathbf{x})\mathbf{x} \tag{118}$$

We may now select the following Liapunov function for this system:

$$\Lambda(\boldsymbol{\delta}) = \tfrac{1}{2}|\boldsymbol{\delta}|^2 = \tfrac{1}{2}\boldsymbol{\delta}^T\boldsymbol{\delta} \tag{119}$$

This obviously meets the first condition for a Liapunov function, that it is positive definite in $\boldsymbol{\delta}$ (which plays the same role as M in section 4.4.). To verify the second condition, we may use equations 118 and 119 to compute:

$$<\partial_t \Lambda> = <\boldsymbol{\delta}^T \partial_t \boldsymbol{\delta}> = \boldsymbol{\delta}^T <\partial_t \boldsymbol{\delta}> = - (\boldsymbol{\delta}^T\mathbf{x})(\boldsymbol{\delta}^T\mathbf{x}) \leq 0 \tag{120}$$

### 7.3. Proofs of Q-Stability of Galerkinized Methods in the Deterministic Case

### 7.3.1. Liapunov Function Common to All Four Systems

For all four systems, I will prove stability using the same Liapunov function:

$$\Lambda(M) = \tfrac{1}{2} \text{Tr}(M^2) \tag{121}$$

Because M is a symmetric real matrix, its square is guaranteed to be a symmetric nonnegative matrix. All of its eigenvalues are guaranteed to be nonnegative. $\Lambda$ is positive definite, because the trace -- the sum of the eigenvalues -- can only be zero if all of the eigenvalues are zero, which would imply M=0. Therefore, equation 121 meets the first condition for a Liapunov function. To complete the proof for each of the four Galerkinized methods to be discussed here, I only need to verify the second condition for each of them.

Because M is a symmetric matrix, we may differentiate equation 121 to get:

$$\partial_t \Lambda(M^2) = \text{Tr}(M(\partial_t M)) \tag{122}$$

### 7.3.1. Second Condition for Q-Stability of HDPGC

The dynamics of M, the error matrix for the weights, in HDPGC are given in equation 66. Substituting that equation into equation 122 (and choosing $\alpha=1$ for simplicity), we get:

$$\delta_t \Lambda = -\text{Tr}(Me(\mathbf{xx}^T A^T + A\mathbf{xx}^T)) = -e(\text{Tr}(M\mathbf{xx}^T A^T) + \text{Tr}(MA\mathbf{xx}^T)) \tag{123}$$

Exploiting the well known theorem (easily proved) that Tr(AB)=Tr(BA), even for nonsquare matrices A and B, we may reduce this to:

$$\partial_t \Lambda = -e(\mathbf{x}^T A^T M\mathbf{x} + \mathbf{x}^T MA\mathbf{x}) \tag{124}$$

Substituting in from equation 64, this is simply:

$$\partial_t \Lambda = -(\mathbf{x}^T A^T M\mathbf{x} + \mathbf{x}^T MA\mathbf{x})^2 \leq 0 \tag{125}$$

This completes the proof of u-stability.

### 7.3.2. Second Condition for Q-Stability of HDPG

The error dynamics for HDPG in the stochastic case are given in equation 84. For the deterministic case, where Q=0 and $\mathbf{w}=\mathbf{0}$, equation 84 reduces to:

$$\partial_t M = d\,(\mathbf{xx}^T - A\mathbf{xx}^T A^T) \,, \tag{126}$$

where I define the following new scalar in order to simplify the algebra:

$$d = \mathbf{x}^T(A^T MA - M)\mathbf{x} \tag{127}$$

Substituting equation 126 into equation 122, we get:

$$\partial_t \Lambda = \text{Tr}(M(\partial_t M)) = d\,\text{Tr}(M\mathbf{xx}^T - MA\mathbf{xx}^T A^T)$$

$$= d\mathbf{x}^T M\mathbf{x} - d\mathbf{x}^T A^T MA\mathbf{x} = d\mathbf{x}^T(M - A^T MA)\mathbf{x} = -d^2 \leq 0 \tag{128}$$

Again, QED.

### 7.3.3. Second Condition for Q-Stability of DHPG

The error dynamics for DHPG in the stochastic case are given in equation 98. For the deterministic case, where Q=0, equation 98 reduces to:

$$\partial_t M = (A^T MA - M)\mathbf{xx}^T - A(A^T MA - M)\mathbf{xx}^T A^T \tag{129}$$

If, for convenience, we define the symmetric matrix:

$$D = A^T M A - M, \quad (130)$$

equation 129 is simplified to:

$$\partial_t M = D\mathbf{x}\mathbf{x}^T - AD\mathbf{x}\mathbf{x}^T A^T \quad (131)$$

Following equation 122, we derive:

$$\partial_t \Lambda = Tr(M(\partial_t M)) = Tr(MD\mathbf{x}\mathbf{x}^T) - Tr(MAD\mathbf{x}\mathbf{x}^T A^T) = \mathbf{x}^T M D\mathbf{x} - \mathbf{x}^T A^T MAD\mathbf{x}$$

$$= \mathbf{x}^T(M - A^T MA)D\mathbf{x} = -\mathbf{x}^T D^2 \mathbf{x} = -|D\mathbf{x}|^2 \le 0 \quad (132)$$

QED.

### 7.3.4. Q-Stability of GDHPG (with $\Omega_0 = 0$ and $\Omega = I$)

The error dynamics for GDHPG in the stochastic case, for $\Omega_0 = 0$ and $\Omega = I$, are given in equation 102. Equation 102 is really just the symmetrized version of equation 98.
In the deterministic case, where Q=0, it reduces to the symmetrized version of equation 129, which can be expressed as follows:

$$M = D\mathbf{x}\mathbf{x}^T - AD\mathbf{x}\mathbf{x}^T A^T + \mathbf{x}\mathbf{x}^T D - A\mathbf{x}\mathbf{x}^T D A^T, \quad (133)$$

using the definition of "D" given in equation 130. Following equation 122, we derive:

$$\partial_t \Lambda = Tr(M(\partial_t M)) = Tr(MD\mathbf{x}\mathbf{x}^T) - Tr(MAD\mathbf{x}\mathbf{x}^T A^T) + Tr(M\mathbf{x}\mathbf{x}^T D) - Tr(MA\mathbf{x}\mathbf{x}^T D A^T)$$

$$= \mathbf{x}^T MD\mathbf{x} - \mathbf{x}^T A^T MAD\mathbf{x} + \mathbf{x}^T DM\mathbf{x} - \mathbf{x}^T D A^T MA\mathbf{x} \quad (134)$$

$$= \mathbf{x}^T(M - A^T MA)D\mathbf{x} + \mathbf{x}^T D(M - A^T MA)\mathbf{x} = -\mathbf{x}^T D^2 \mathbf{x} - \mathbf{x}^T D^2 \mathbf{x} = -2|D\mathbf{x}|^2 \le 0$$

### 7.4. Mathematical Preliminaries to Proofs of Q-Instability

Once we have a Liapunov function in hand, it can be very straightforward to prove q-stability, as you can see in section 7.3. For a given plant, instability can also be proven in a straightforward way, by giving examples of training schedules {$\mathbf{x}$(t)} which cause an explosion in the estimation errors M. However, to prove q-instability across a broad range of possible plants is more difficult. It requires a more abstract mathematical approach, even though the goal here is simply to develop a better practical understanding of how these systems work. This subsection will discuss some of the required mathematics.

First of all, our focus here will be on q-instability for four methods -- HDPC, HDP, DHP and GDHP -- in the deterministic linear case. This reduces down to an analysis of four different dynamical systems for the matrix M -- equation 65 (for HDPC),
equation 73 (for HDP), equation 105 (with angle brackets removed, for generalized DHP) and equation 106 (likewise, for generalized GDHP). It can easily be seen that these dynamical systems are linear in the matrix M. Thus the easiest way to analyze these systems is by applying linear systems theory to them. In this case, it requires a certain amount of care to do this.

Our original dynamical system (equations 44 or 45) involved a state vector, $\mathbf{x}$, with n components. In formal terms, $\mathbf{x} \in R^n$. But the linear deterministic systems of equations 65,7 73, 105 and 106 concern a matrix M. In order to apply ordinary linear systems theory, we may think of M, in equation 105, as a kind of vector of dimensionality $N=n^2$. In other words, $\underline{\mathbf{M}} \in R^N$ is the representation of the matrix M as a vector in $n^2$-dimensional space. For equations 65, 73 and 106, M is taken from the space of <u>symmetric</u> matrices, which only have $N'=n(n+1)/2$ independent degrees of freedom; then in those cases, the vector representation of M will be $\underline{\mathbf{M}} \in R^{N'}$. Throughout this section, I will define the quantities n, N' and N in this way.

For q-stability in the linear case, we required that $\Lambda(C,Q,V,A)$ be quadratic in C. The Liapunov function can be expressed as a quadratic about the desired stable equilibrium, C*; in other words, it can be expressed as a quadratic in M, if such a Liapunov function exists. Because Q=0 in the linear deterministic case, and because the dynamics of M in equations 65, 73, 105 and 106 do not depend upon V, this means that we require $\Lambda(M,A)$ to be quadratic in M. Thus we are looking for a Liapunov function of the form:

$$\Lambda(M,A) = \tfrac{1}{2} \sum_{ijkl} T^{ijkl}(A) M_{ij} M_{kl} \qquad (135)$$

Even though $T^{ijkl}$ should always be understood to be a function of A, I will sometimes leave this dependence implicit below.

Without loss of generality, we can require that T be symmetric in M; in other words:

$$T^{ijkl} = T^{klij} , \qquad \text{for all } i,j,k,l \qquad (136)$$

Furthermore, if we think of $\underline{\mathbf{M}}$ as a vector in $R^N$ or $R^{N'}$, we may also define:

$$\underline{\mathbf{L}} = \nabla \Lambda = \sum_{i,j} T^{ijkl}(A) M_{ij} , \qquad (137)$$

where $\underline{\mathbf{L}}$ also corresponds to a matrix L. In the case where $\underline{\mathbf{M}}$ is a vector in $R^{N'}$, we may also require, without loss of generality, that:

$$T^{ijkl} = T^{jikl} = T^{ijlk} = T^{jilk} , \qquad (138),$$

and that L is therefore a symmetric matrix.

The first condition for q-stability requires that T(A) must be positive definite, when considered as an N-by-N or N'-by-N' matrix. The second condition requires that:

$$\partial_t \Lambda = \sum_{ijkl} T^{ijkl} M_{ij}(\partial_t M_{kl}) = \mathrm{Tr}(L^T \partial_t M) \le 0 \qquad (139)$$

for all M and $\underline{\mathbf{x}}$, for the particular matrix A under study. Note that equation 139 specifically follows for the choice of T as a symmetric matrix. It is not possible to satisfy the requirement for q-stability with a matrix T which is not symmetric, unless it is also possible to satisfy it for the symmetrized version of that matrix. Thus to prove q-instability in general, it is enough to prove that no such symmetric T exists.

### 7.5. Q-Instability of HDPC

Could the requirement in equation 138 be satisfied for some symmetric T, in HDPC?
To evaluate this, substitute equation 65 into equation 138 (with $\alpha=\frac{1}{2}$), which yields the requirement:

$$\text{Tr}(L(\tfrac{1}{2}e\underline{x}\underline{x}^T)) \leq 0 \quad \text{for all } \underline{x}, M \tag{140}$$

Note that L is a symmetric matrix, in this case, because M is a symmetric matrix, and because equation 138 is imposed.
Substituting equation 64 into 140, the requirement of equation 138 reduces to:

$$(\underline{x}^T L \underline{x})(\underline{x}^T(A^T M + MA)\underline{x}) \leq 0 \quad \text{for all } \underline{x} \text{ and } M \tag{141}$$

The analysis of equation 141 leads to the following lemma:

**7.5.1. Lemma on L and B**: If:

$$(\underline{x}^T L \underline{x})(\underline{x}^T B \underline{x}) \leq 0 \text{ for all } \underline{x} \tag{142}$$

for some symmetric matrices L and B, then either: (1) B is positive semidefinite and L is negative semidefinite; (2) L is positive semidefinite and B is positive semidefinite; or (3) L=-kB for some nonnegative scalar k.

Proof: Since L is real and symmetric, we may, without loss of generality, assume that a similarity transformation is performed on L, B and $\underline{x}$ based on the eigenvectors of L, such that L becomes a diagonal matrix. After that, if the elements on the diagonal of L are all nonnegative or all nonpositive, then it is obvious that $\underline{x}^T L \underline{x}$ is always negative or always positive; in those cases, equation 142 requires that $\underline{x}^T B \underline{x}$ is always nonpositive or always nonnegative, respectively, for all $\underline{x}$, which then gives us cases (2) and (1) respectively.
The only alternative possibility is that L contains a mix of diagonal elements (eigenvalues), some positive and some negative.
Let $\underline{e}_i$ and $\underline{e}_j$ be the basis vectors (after the similarity transformation) corresponding to one of the positive eigenvalues $\lambda_i$ of L, and one of the negative eigenvalues, $\lambda_j$, respectively.
If we consider the case $\underline{x}=\underline{e}_i$, equation 142 tells us that $B_{ii} \leq 0$. Likewise, the case $\underline{x}=\underline{e}_j$ tells us that $B_{jj} \geq 0$. For $\underline{x}=\underline{e}_i+y\underline{e}_j$, we may calculate:

$$\underline{x}^T L \underline{x} = \lambda_i + y^2 \lambda_j \tag{143a}$$

$$\underline{x}^T B \underline{x} = B_{ii} + 2yB_{ij} + y^2 B_{jj} \tag{143b}$$

The graph for equation 143a as a function of y is a parabola, crossing the x-axis at two points, where $y = \pm \sqrt{-\lambda_i/\lambda_j}$. Equation 142 tells us that the quadratic expression in 143b must cross the x-axis at the same two points. From elementary algebra, this tells us that $B_{ij}=0$ and $B_{ii}/B_{jj} = \lambda_i/\lambda_j$.
If we then consider this same argument, applied to all possible positive/negative $\lambda_i/\lambda_j$ pairs (including $\lambda_k/\lambda_j$ involving different k but the same j), we may deduce that the entire diagonal of B must equal -k multiplied by L, for some nonnegative scalar k. We may also deduce that all the $B_{ij}$ crossterms for $\lambda_i$ positive and $\lambda_j$ negative must be zero.
Next, consider $\underline{x}$ of the form $\underline{e}_i + y\underline{e}_j + z\underline{e}_k$, where $\lambda_k, \lambda_i > 0$ and $\lambda_j < 0$. We again arrive at two expressions quadratic in y, analogous to 143a and 143b, which must match crossing points as before, for every possible value of z; the only way this can happen is if $B_{ik}=0$. In a similar way, we may deduce that the offdiagonal terms in the negative block of B are also zero. In summary, we have deduced that <u>all</u> of the offdiagonal terms of B must be zero -- the ones which cross between the negative and positive blocks, and the ones which go between different elements within each block. Thus B must be a diagonal matrix. Since we have already shown that the diagonal of B must equal -kL, this completes the proof of the lemma.

### 7.5.2. HDPC: Derivation of the Required Form of L

Applying the Lemma of section 7.5.1 to equation 141, we may deduce the requirement that:

$$L(A,M) = -k(A,M)(A^T M + MA) , \qquad (144)$$

for all matrices M such that $A^T M+MA$ is not positive semidefinite or negative semidefinite.

Let us say that $M \in P(A)$ when $A^T M+MA$ is positive semidefinite. It is straightforward to prove that P(A) is a convex cone[81] by simply applying the definition of a convex cone: (1) if $M \in P(A)$, then $kM \in P(A)$ (for $k \geq 0$) because k multiplied by a positive semidefinite matrix is still a positive semidefinite matrix; (2) if $M_1 \in P(A)$ and $M_2 \in P(A)$, then
$(1-k)M_1+kM_2$ (for $1>k>0$) is in P(A), because the sum of two positive definite matrices is also positive definite. Let us say that $M \in N(A)$ iff $-M \in P(A)$; this is also a closed convex cone. Let us define K(A) as the rest of $R^n$, after P(A) and N(A) are removed; clearly it, too, is a connected cone.

Equation 137 tells us that L would be required to be linear in M. Consider any two matrices, $M_1$ and $M_2$, both in K(A), such that $M_1+M_2$ is also in K(A), and that $A^T M_1+M_1 A$ is not just equal to a scalar multiple of $A^T M_2+M_2 A$. In this case, the linearity of L as a function of M requires that $k(A,M_1)=k(A,M_2)=k(A,M_1+M_2)$. Likewise, the linearity of L requires that $k(A,cM)=k(A,M)$ for any scalar c. Since all matrices M in K(A) are connected to each other by some combination of adding and scaling, this tells us that k(A,M) is actually k(A) -- a constant with respect to M, all across K(A). Furthermore, all matrices in P(A) or N(A) can be reached as linear combinations of matrices in K(A). (In other words, it is easy to write any positive definite symmetric matrix as a <u>linear combination</u> of two symmetric matrices which are not definite; at this stage, we do not require a convex combination, only a linear combination.) Therefore, we may deduce:

$$L(A,M) = -k(A)(A^T M+MA) \quad \text{for all M,} \qquad (145)$$

for <u>any linear dynamic process (A) which is q-stable</u>.

### 7.5.3. HDPC: Further Analysis of L and Conclusions

Combining equations 137 and 145, we obtain the requirement that:

$$\sum_{kl} T^{ijkl}(A)M_{kl} = -k(A) \sum_k (A_{ik}^T M_{kj} + M_{ik}A_{kj}) \quad \text{for all M} \qquad (146)$$

Differentiating with respect to $M_{kl}$ (recalling that $M_{kl}=M_{lk}$ is <u>one element</u> of $\mathbf{M} \in R^{N'}$) and exploiting the fact that we can multiply T by a positive scalar without changing its validity as a Liapunov function, we may deduce the requirement that:

$$T^{ijkl}(A) = -(\delta_{jl}A_{ik}{}^T + \delta_{jk}A_{il}{}^T + \delta_{il}A_{kj} + \delta_{ik}A_{lj}) \qquad (147)$$

Again, equation 147 has been deduced as a <u>necessary</u> requirement for a symmetric, quadratic Liapunov function T for a deterministic linear system. But equation 136 is another necessary requirement. The combination of these two requirements, which apply to <u>all</u> combinations of i,j,k and l, implies a requirement that:

$$\delta_{jl}A_{ik}{}^T + \delta_{jk}A_{il}{}^T + \delta_{il}A_{kj} + \delta_{ik}A_{lj} = \delta_{lj}A_{ki}{}^T + \delta_{li}A_{kj}{}^T + \delta_{kj}A_{il} + \delta_{ki}A_{jl} \qquad (148)$$

For example, in the case where i=j=k, this requires that:

$$\delta_{il}A_{ii}{}^T + A_{il}{}^T + \delta_{il}A_{ii} + A_{li} = \delta_{li}A_{ii}{}^T + \delta_{li}A_{ii}{}^T + A_{il} + A_{il} \qquad (149)$$

which reduces to:

$$2A_{li} = 2A_{il} \quad \text{for all i and l} \qquad (150)$$

which is satisfied if and only if the matrix A is symmetric.

The conclusion of this analysis is that HDPC can be q-stable only in the case where A happens to be symmetric. Matrices of this sort are a set of measure zero in the space of al matrices A! More importantly, there are few real-world applications where the dynamic matrix A is exactly symmetric. This is far more restrictive than the restrictions cited in section 3 for the old IAC designs. When A <u>does</u> happen to be symmetric, it can be shown that equation 147 does give us a valid Liapunov function; however, since HDPC clearly fails to meet our initial goals in this paper, I will not elaborate further on that point.

Notice that L, the gradient of Λ, is defined for any differentiable Liapunov function over M. Thus equations 140, 141 and 144 are valid requirements for unconditional stability in general, not just q-stability. Q-stability was used in this argument only to derive the more narrow requirement given in equation 145. Prior to the development of this proof, I analyzed a variety of two-by-two and diagonalized (complex) examples, where the results were all consistent with the general result given here. Intuitively, the argument here is saying that the required gradient field L is not an integrable field, because it does not meet the usual
condition for the integrability of a vector field; in other words, the required L cannot actually be the gradient of a scalar field Λ, as required, because its derivatives do not possess the required symmetry.

Equation 144 seems very close to equation 145, on the surface; however, section 8 will explain why this may be misleading. I would now conjecture that there are some dynamical systems A, which are not <u>exactly</u> symmetric, where HDPC is unconditionally stable, based on a Liapunov function which is not quadratic. However, this does not imply unconditional stability across <u>all</u> controllable systems A, which is what we really want.

### 7.6. Q-Instability of HDP Proper

The proof of q-instability of HDP proper follows exactly the same structure as the proof in section 7.5. After all, the dynamics of M in equation 73 are virtually identical to the dynamics in equation 65, except that we have a different definition of "e" (from equation 72 instead of 64).

More precisely, the proof of q-instability for HDP is exactly the same as in section 7.5, up to the start of section 7.5.3, except that "A$^T$M+MA" is replaced everywhere by "A$^T$MA-M". In place of equation 146, we then have the requirement that:

$$\sum_{kl} T^{ijkl} M_{kl} = -k(A)(-M_{ij} + \sum_{kl} A_{ik}^T M_{kl} A_{lj}) \qquad (151)$$

for all M. Once again, we may differentiate with respect to $M_{kl}=M_{lk}$ and exploit the invariance of T with respect to positive scalar multiplication, to derive the requirement:

$$T^{ijkl} = \delta_{ik}\delta_{jl} + \delta_{il}\delta_{jk} - (A_{ik}^T A_{lj} + A_{il}^T A_{kj}) \qquad (152)$$

Once again, equation 134 then requires:

$$\delta_{ik}\delta_{jl} + \delta_{il}\delta_{jk} - A_{ik}^T A_{lj} - A_{il}^T A_{kj} = \delta_{ki}\delta_{lj} + \delta_{kj}\delta_{li} - A_{ki}^T A_{jl} - A_{kj}^T A_{il} \qquad (153)$$

for all combinations of i, j, k, and l. For example, if we chose k=l, and sum the results over k, equation 15e reduces to the necessary requirement that:

$$-A^T A - A^T A = -AA^T - AA^T \qquad (154)$$

This tells us that A commutes with $A^T$, which tells us that both of them commute with $A+A^T$. This tells us that A is a "normal matrix"[81,82], and that its eigenvectors (like those of $A+A^T$) will usually be real and orthogonal. (Exceptions may be possible if $A+A^T$ has degenerate eigenvalues.) When a real matrix has eigenvectors which are all real and orthogonal, it must be symmetric. Once again, the matrix A must either be symmetric, or it must fall into some other set of exceptions which is also of measure zero.

The requirement given in equation 154 is only a necessary requirement for q-stability, not a sufficient requirement. As in section 7.5, however, the necessary requirement is already strong enough to tell us that HDP does not provide q-stability for a broad enough
range of dynamical systems.

**7.7. Q-Instability of GDHP**

Because GDHP preserves the symmetry of M, while DHP does not, it is easier to begin by considering GDHP.

For GDHP, q-stability again requires that we find a quadratic Liapunov function which satisfies equations 136 through 139. In the deterministic case, the error dynamics fro GDHP as given in equation 106 may be written as:

$$\partial_t M = \Omega(A^T M A - M)\underline{xx}^T + \underline{xx}^T(A^T M A - M)\Omega = D\underline{xx}^T + \underline{xx}^T D^T, \qquad (155)$$

where I define:

$$D = \Omega(A^T M A - M) \qquad (156)$$

Substituting equation 155 into equation 139, and recalling that L must be symmetric in this case, we deduce the requirement that:

$$\text{Tr}(LD\underline{xx}^T + L\underline{xx}^T D^T) = \underline{x}^T LD\underline{x} + \underline{x}^T D^T L\underline{x} = 2(L\underline{x})^T(D\underline{x}) \leq 0 \qquad (157)$$

for all vectors **x**.

It is safe at this point to assume that D and L are both nonsingular, for the following reasons. First, we may set aside the case A=±I. Our main goal here is to show that GDHP is almost always q-unstable, when A is <u>not</u> symmetric. Second, for a sensible adaptation algorithm, we require that Ω be nonsingular. Third, we have required that Λ be positive definite, which implies that L must be nonsingular when M is. In summary, for all nonsingular M, L and D must be nonsingular (for q-stability), which would allow us to define the matrix:

$$D = S(A,M)L \tag{158}$$

Inserting this definition into equation 157 tells us that S(A,M) must be a negative semidefinite matrix, for all M and A. (By continuity, we expect this for <u>all</u> matrices M.)

Combining equation 158 with equations 137 and 156, we deduce the requirement for q-stability:

$$L_{ij} = \sum_{kl} T^{ijkl}M_{kl} = (S^{-1}\Omega(A^T MA - M))_{ij} \tag{159}$$

The analysis of equation 159 leads to the following lemma:

### 7.7.1. Lemma of Tensor Linearity. Suppose that:

$$\sum_{kl} P^{ijkl}M_{kl} = \sum_{kl\nu} B^{i,kl}_{\nu}M_{kl}C_{\nu j}(M) \tag{160}$$

for all symmetric matrices M, for some tensors P and B which are not functions of M. Let us also assume that B has the property that B**M** is nonsingular almost everywhere, in the space of **M**∈$R^{N'}$. Then $C_{\nu j}(M)=C_{\nu j}(0)$ is not actually a function of M.

Proof: The validity of equation 160 implies the validity of equation 160 differentiated with respect to $M_{\alpha\beta}$, for any α and β; this yields:

$$A^{ij}_{\alpha\beta} = \sum_{\nu} B^{i,\alpha\beta}_{\nu}C_{\nu j}(M) + \sum_{kl\nu} B^{i,kl}_{\nu}M_{kl}\frac{\partial C_{\nu j}}{\partial M_{\alpha\beta}} \tag{161}$$

If we multiply by $M_{\alpha\beta}$ and sum over α and β, we deduce the requirement that:

$$\sum_{\alpha\beta} A^{ij}_{\alpha\beta}M_{\alpha\beta} = \sum_{\alpha\beta\nu} B^{i,\alpha\beta}_{\nu}M_{\alpha\beta}C_{\nu j}(M) + \sum_{kl\nu\alpha\beta} B^{i,kl}_{\nu}M_{\alpha\beta}M_{kl}\frac{\partial C_{\nu j}}{\partial M_{\alpha\beta}} \tag{162}$$

But the first two terms in equation 162 are essentially identical to equation 160. Subtracting out equation 160 from equation 162, we deduce:

$$\sum_{kl\nu\alpha\beta} B^{i,kl}_{\nu} M_{\alpha\beta} M_{kl} \frac{\partial C_{\nu j}}{\partial M_{\alpha\beta}} = 0 \tag{163}$$

Let us define the matrices:

$$D_{i\nu} = \sum_{kl} B^{i,kl}_{\nu} M_{kl} \tag{164a}$$

$$R_{\nu j} = \sum_{\alpha\beta} M_{\alpha\beta} \frac{\partial C_{\nu j}}{\partial M_{\alpha\beta}} \tag{164b}$$

Using these definitions, equation 103 may be expressed as:

$$D(M)R(M) = 0 \tag{165}$$

Because B**M** is assumed to be nonsingular almost everywhere in M, D(M) will be nonsingular almost everywhere. Thus almost everywhere in M, equation 165 requires:

$$R(M) = 0 \tag{166}$$

By the requirement for continuity with respect to M, R(M)=0 is required <u>for all M</u>. Equation 166, plus the definition in equation 164b, are an instance, in $R^{N'}$, of the equations:

$$\underline{x}^T \nabla f(\underline{x}) = 0 \text{ for all } \underline{x} \tag{167}$$

Integrating equation 167 along any line radiating out from the origin to a point $\underline{x}$, we may deduce in general that it implies $f(\underline{x})=f(\underline{0})$. IN other words, we may deduce that f is actually a constant with respect to $\underline{x}$. Applying this to equations 164b and 166, we may deduce that $C_{\nu j}(M)=C_{\nu j}(0)$ is not actually a function of M.

### 7.7.2. Derivation of the Form of L

Applying equation 160 and the lemma in section 7.7.1 to the analysis of equation 159, we may deduce that $S^{-1}$ cannot actually be a function of M. Strictly speaking, however, we need to verify that B**M** will be nonsingular almost everywhere in this case. In this case, B**M** is simply the right-hand side of equation 159, which, for q-stability, must equal L. But in the discussion leading up to equation 158, we already explained why we require that L be nonsingular almost everywhere in M.

Once we accept that $S^{-1}$ cannot be a function of M, we may proceed as follows. First define the matrix:

$$R(A) = S^{-1}(A)\Omega \tag{168}$$

Inserting this definition into equation 159, we get:

$$L = R(A)(A^T M A - M) \quad , \tag{169}$$

but L is required to be symmetric <u>for all symmetric matrices M</u>.
Assuming that the N'-by-N' matrix which multiplies **M** to generate <u>$A^T M A - M$</u> is not singular, this implies that $A^T M A - M$ ranges over all possible symmetric matrices as well.
(Again, I exclude consideration of the case A=±I.) If R(A)D is symmetric for <u>all</u> symmetric matrices D, then R(A) must be of the form kI, where k is a scalar. Substituting this back into equation 169, and recalling that T is only defined up to a positive scalar anyway (and using some knowledge about systems analysis to set the sign), we deduce the requirement that:

$$\sum_{kl} T^{ijkl} M_{kl} = M_{ij} - \sum_{kl} A_{ik}^T M_{kl} A_{lj} \tag{170}$$

Differentiating with respect to $M_{kl} = M_{lk}$, we obtain the requirement:

$$T^{ijkl} = \delta_{ik}\delta_{jl} - A_{ik}^T A_{lj} + \delta_{il}\delta_{jk} - A_{il}^T A_{kj} \tag{171}$$

Once again, we invoke equation 136 to deduce the requirement:

$$\delta_{ik}\delta_{jl} - A_{ik}^T A_{lj} + \delta_{il}\delta_{jk} - A_{il}^T A_{kj} = \delta_{ki}\delta_{lj} - A_{ki}^T A_{jl} + \delta_{kj}\delta_{li} - A_{kj}^T A_{il} \tag{172}$$

for all i, j, k and l. This reduces to:

$$A_{ik}^T A_{lj} + A_{il}^T A_{kj} = A_{ki}^T A_{jl} + A_{kj}^T A_{il} \tag{173}$$

Consider the case k=l and sum over l to deduce the necessary requirement:

$$A^T A + A^T A = A A^T + A A^T \tag{174}$$

This is essentially the same as equation 154. As in section 7.6, it tells us that q-stability requires that A must either be symmetric, or must belong to some other set of exceptional cases of measure zero. Thus GDHP, like HDPC and HDP proper, is q-unstable for almost all matrices A, under the conditions of this paper.

### 7.8. Q-Instability of Generalized DHP

For q-stability of DHP, we would still require equations 135-137 and 139 to hold; however M and L are no longer required to be symmetric, and equation 138 does not apply.
    When we substitute the adaptation rule for generalized DHP in the deterministic case (equation 105 with angle brackets removed) into equation 139, we deduce the requirement for q-stability that:

$$Tr(L^T \partial_t M) = Tr(L^T D \mathbf{x}\mathbf{x}^T) = \mathbf{x}^T L^T D \mathbf{x} = (L\mathbf{x})^T (D\mathbf{x}) \leq 0 \tag{175}$$

where D is again defined as in equation 156. Note the similarity of the final inequality in equation 175 to the final inequality in equation 157. Starting from that inequality, we may use the exact same logic as before, in order to deduce that equations 158 and 159 are also required here. Likewise, the lemma of section 7.7.1 can be applied. Thus we may deduce that q-stability for DHP will require that equation 159 hold, for $S^{-1}$ <u>not</u> a function of M. This again allows the definition in equation 168, which leads to equation 169.

For DHP, however, we <u>cannot</u> deduce that R(A) must equal kI. The requirement for q-stability given in equation 169 may be written out more exhaustively and combined with equation 137, to yield the requirement:

$$L_{ij} = \sum_{ij\nu} R_{k\nu} A_{\nu i}^T M_{ij} A_{jl} - \sum_{\nu} R_{k\nu} M_{\nu l} = \sum_{ij} T^{ijkl} M_{ij} \quad \text{for all M} \quad (176)$$

Differentiating with respect to $M_{ij}$, we obtain the requirement:

$$\sum_{\nu} R_{k\nu} A_{\nu i}^T A_{jl} - R_{ki} \delta_{lj} = T^{ijkl} \quad (177)$$

Applying equation 136, we may deduce the requirement, for all i, j, k and l, that:

$$(RA^T)_{ki} A_{jl} - R_{ki} \delta_{lj} = (RA^T)_{ik} A_{lj} - R_{ik} \delta_{jl} \quad (178)$$

Consider the case where l≠j. If all of the crossterms $A_{lj}$ for l≠j are zero, the A would be a symmetric matrix. But our goal here is to show q-instability for almost all A, <u>except</u> for the case of A symmetric. If a non-symmetric matrix A is q-stable, then non-symmetry would imply that $A_{lj} \neq 0$ for some j≠l. For all such (j,l) pairs, two of the terms in equation 178 drop out, and we may divide what remains by $A_{lj}$ to get:

$$(A_{jl}/A_{lj})(RA^T)_{ki} = (RA^T)_{ik} \quad (179)$$

for all i and k. In order to satisfy this equation with $A_{jl}=0$, for all i and k, it would be necessary that $(RA^T)_{ik}=0$ for all i and k; this would require that R be singular (which, applied to equation 176, would violate our requirement that Λ be positive definite) or that A=0 (which is a symmetric case). Thus we may deduce that $A_{jl} \neq 0$, and that we may find at least some combination of k and i such that $(RA^T)_{ki} \neq 0$. For every such combination of k and i, we may divide equation 179 to deduce:

$$A_{jl}/A_{lj} = (RA^T)_{ik}/(RA^T)_{ki} \quad (180)$$

We may also deduce that the numerators and denominators of equation 180 are all nonzero, for these combinations (l,j) and (k,i). Because they must all be nonzero, we can apply the same logic for l and j interchanged, or for k and i interchanged, independently.
This implies, for example, that:

$$A_{jl}/A_{lj} = A_{lj}/A_{jl} = \pm 1 \quad (181)$$

Furthermore, since we may use the exact same logic for <u>any</u> nonzero $A_{\mu\nu}$ (with μ≠ν), using the <u>same</u> combination of k and i, we may deduce the following. For q-stability, for any matrix A which is not symmetric, <u>every</u> offdiagonal term would have to obey $A_{lj} = -A_{jl}$.
In that case, equation 180 would also require that $RA^T$ would have to be completely antisymmetric.

This analysis could be extended further, to provide additional restrictions or requirements for q-stability. However, we have already shown that q-stability is possible here only in two situations -- where A is completely symmetric, or where the offdiagonal terms are completely antisymmetric. Both situations represent sets of measure zero in the space of possible matrices A. Therefore, it is clear that DHP does not fulfill the goals of this paper,
and more than HDPC, HDP or GDHP.

## 8. Interpretation of the Results in Sections 6 and 7

What are the long-term strategic implications of the results in sections 6 and 7? What do they tell us about long-term opportunities for the program discussed in section 5?

At first glance, the results seem to be very discouraging. We do know that real-time intelligent systems, as discussed in section 5.3, are possible; after all, biological organisms do exist. But this does not imply that unconditional stability is a realistic goal; after all, the experience of biological organisms is not a "movie constructed by a devil" (as discussed in section 7.1). This suggests the minimal strategy of continuing to use established Critic adaptation methods, and proving theorems which exploit the requirement that the schedule of experience **x**(t) should come from an appropriate probability distribution [73].

This section will discuss some alternative interpretations of the problem, which suggest possible ways to overcome it. Section 8.1 will discuss the gap between q-stability and unconditional stability. This gap is significant, but it seems to offer little hope that the established Critic adaptation methods could be unconditionally stable across the majority of controller plants A.

Section 8.2 will discuss the Generalized Moving Target (GMT) problem previously discussed in [17]. It will offer new gradient-based neural network designs to address that problem. It will also address related possibilities for assuring convergence more reliably (but still cheaply) for large-scale economic equilibrium models like the one discussed in [16]. However, it will also show how the application of these designs to the Critic adaptation problem here involve excessive cost, just like the previous stable methods of Landelius [43]
and Bradtke [44].

Finally, section 8.3 will discuss a more intuitive interpretation of the problem which, though simple, leads naturally to the construction of new methods, to be given in section 9.

### 8.1. Q-Stability Versus Unconditional Stability

The dynamic equations for M given in section 4.4 are all examples of a larger class of systems called Linear Differential Inclusions (LDI) [79]. In essence, we are considering
a special case of a larger problem: for the system

$$\partial_t \mathbf{x} = A(\alpha)\mathbf{x}, \qquad (182)$$

starting from some bounded initial state **x**(0), prove that **x**(t) at later times is also bounded, regardless of the choice of $\alpha(t)$ at later times t. Equation 182 is not a linear dynamic system, because it refers to a <u>family</u> of matrices $A(\alpha)$, for different possible values of the parameter(s) $\alpha$, rather than a single matrix A. In our application, the state vector being studied is actually the matrix **M**, and the set of parameters $\alpha$ is actually the state vector of the plant being controlled.

Boyd et al[79, p.74] cite many examples of systems of this sort which are unconditionally stable, but not q-stable. However, these examples do not give us a complete intuitive feel for the nature of the gap between q-stability and unconditional stability in our situation. I will present another example here, in a purely intuitive fashion, in order to help build up this kind of intuitive feeling.

Suppose that $\underline{x}$ is simply a two-component vector, $(x_1, x_2)$. Suppose that the parameter $\alpha$ takes on only two possible values, $\alpha=1$ and $\alpha=2$, with:

$$A(1) = \begin{bmatrix} a & 1 \\ -1 & b \end{bmatrix}; \qquad A(2) = \begin{bmatrix} b & 1 \\ -1 & a \end{bmatrix} \qquad (183)$$

where a>0 and b<0 are both very small (near zero). In order to determine whether this system can be destabilized, we need to consider the control problem of how to pick $\alpha(\underline{x})$ in order to maximize future $|\underline{x}|^2$ over future times t. In this case, the choice between $\alpha=1$ and $\alpha=2$ is easy because:

$$\partial_t (\tfrac{1}{2}|\underline{x}|^2) = \underline{x}^T (\partial_t \underline{x}) = ax_1^2 + bx_2^2 \text{ (for } \alpha=1\text{)} \text{ or } bx_1^2 + ax_2^2 \text{ (for } \alpha=2\text{)} \qquad (184)$$

Thus at each point $\underline{x}$, we get more increase (or less decrease) in length by choosing A(1) when $|x_1|>|x_2|$ and A(2) when $|x_2|>|x_1|$. If a is very small, and we set b=-ka, and consider $\underline{x}$ on the unit circle (which we can do without loss of generality here), we may integrate equation 184 over the interval from $\theta=-\pi/4$ to $\theta=+\pi/4$, in order to evaluate the total net increase or decrease in the length of $\underline{x}$ over that quarter circle. (The other quarter circles would involve the same increase or decrease, by the symmetry here.) In order to ensure a decrease of the length of $\underline{x}$ (i.e., in order to ensure stability), we require:

$$k \geq \left( \int_0^{\pi/4} \cos^2\theta\, d\theta \right) / \left( \int_0^{\pi/4} \sin^2\theta\, d\theta \right) = (\pi+2)/(\pi-2) \qquad (185)$$

Thus we need something like k≥5 to ensure stability.

It is easy to see that this system is not q-stable for any choice of k. By symmetry, we would expect $\Lambda=|\underline{x}|^2$ to be the only serious possibility for a quadratic Liapunov function here. But at $\theta = \pm\pi/4$, etc., both of the matrices A(1) and A(2) lead to increases in the length of $\underline{x}$ !

Proving all of this would take further space in this already long paper. But the goal here is only to establish some intuition.

This example shows how we may have a kind of negative boundary zone between A(1) and A(2), which prevents the existence of a global quadratic Liapunov function. Nevertheless, if the dynamics of the system are guaranteed to move the system <u>out</u> of this boundary zone, then there may be no possibility of instability in the overall system.

On the other hand, this example shows that a small destabilizing term like a (in $A_{11}(1)$) requires a much larger counterterm (like -5a) elsewhere, in order to permit stability.

In a similar vein, for methods like HDP, consider the possibility of a dynamical system:

$$A(1) = \begin{bmatrix} -1 & 0 \\ a & -1 \end{bmatrix} \qquad (186)$$

where once again a is very small. Because A is not symmetric, HDP is not q-stable in adapting to this system. An algebraic analysis does show that there are training vectors like $(1, \pm\tfrac{1}{2}a)$ which can destabilize it, for some matrices M. However, continued training in such special cases would be expected to <u>change M</u>. In effect, there are negative boundary zones here as well, where M can grow, but only for a short time. Intuition suggests that HDP might well be unconditionally stable, in fact, for such a system A, even though it is not q-stable.

Nevertheless, as with equation 184, I would expect these mechanisms to work only for plants A which are <u>close to</u> q-stable plants (symmetric A) in some sense. This mechanism does not seem likely to prove that HDP is a universal stable adaptation rule.

Consideration of this example does give some clue as to why large interest rates may tend to stabilize HDP in some situations, as discussed in section 4.3.4. Interest rates add a negative term to the diagonal of A, in effect. This cannot make a nonsymmetric matrix symmetric! However, it can make the matrix <u>closer</u> to a symmetric matrix, insofar as the nonsymmetric offdiagonal terms may become smaller <u>relative to</u> the diagonal terms.

## 8.2. Generalized Moving Target (GMT) Problems and Algebraic Systems

The Generalized Moving Target (GMT) problem may be defined as follows. Find the vectors $\underline{v}$ and $\underline{w}$ which simultaneously meet the two following conditions:

(1) $E(\underline{v}, \underline{w})$ is minimized as a function of $\underline{v}$, for $\underline{w}$ held fixed;

(2) $\underline{w} = \underline{v}$.

In 1990 [17], I pointed out that the equilibrium of HDP solves that problem, locally (meeting the first-order conditions for a minimum), for the error function:

$$E(\underline{v}, \underline{w}) = \sum_t (\hat{J}(\underline{R}(t),\underline{v}) - (\hat{J}(\underline{R}(t+1),\underline{w})+U))^2 \qquad (187)$$

I also mentioned consideration of special-purpose quasiNewton approaches, which may yet be of some value in some applications of this sort.

Our problem with stability, using HDP, may be explained as the result of minimizing E with respect to $\underline{v}$, while not <u>anticipating</u> the effect that this has on changing the problem by changing $\underline{w}$. But minimizing $E(\underline{v},\underline{v})$ directly, as in the Galerkinized methods, would solve the wrong problem. Likewise, all manner of Bayesian updates rooted in the effort to minimize $E(\underline{v},\underline{v})$ would be nothing but sophisticated ways to solve the wrong problem, to move to an incorrect equilibrium. What can we do?

One obvious approach is to try to solve the GMT problem in general, and then apply the result to the adaptation of Critics. In any event, this approach can help us to better understand the nature of our problems in section 7.

### 8.2.1. Designs to Solve the GMT Problem

As a preliminary to finding a general solution, consider the quadratic special case of GMT, QGMT, where we face:

$$E(\underline{v}, \underline{w}) = \underline{v}^T A \underline{v} + \underline{w}^T B \underline{w} + \underline{v}^T C \underline{w} + \underline{x}^T \underline{v} + \underline{y}^T \underline{w}, \qquad (188)$$

where A, B, C, $\underline{x}$ and $\underline{y}$ are arrays of parameters held fixed, The first-order condition for a minimum of equation 188 with respect to each component $v_i$ of $\underline{v}$ is:

$$E(\underline{v}, \underline{w}) = \sum 2A_{ij}v_j + \sum C_{ij}w_j + x_i = 0 \qquad (189)$$

j             j

or simply:

$$\nabla_{\underline{v}} E(\underline{v}, \underline{w}) = 2A\underline{v} + C\underline{w} + \underline{x} = \underline{0} \tag{190}$$

Following the gradient of E, as in gradient-based HDP, we would arrive at an adaptation rule:

$$\partial_t \underline{v} = -(2A\underline{v} + C\underline{w} + \underline{x}) \tag{191}$$

Combined with the obvious companion adaptation rule, that $\underline{w}=\underline{v}$, this leads to the overall dynamics:

$$\partial_t \underline{v} = -(2A+C)\underline{v} - \underline{x} \tag{192}$$

When this process reaches equilibrium, we are guaranteed that it satisfies equation 190. Like gradient-based HDP, it does converge to the correct equilibrium -- when it converges at all. However, equation 192 will converge only when 2A+C meets the usual Hurwicz conditions, of having eigenvalues all in the left-hand side of the complex plane. But the actual solution of the GMT problem is well-defined whenever the symmetric matrix A is negative definite.
There are many situations where this condition is met, but 2A+C would not give convergence, How can we achieve reliable convergence for <u>all</u> instances of the problem where the solution is well-defined? The <u>total</u> gradient -- the gradient of $E(\underline{v}, \underline{v})$ -- is almost useless here, since it drags in the matrix B, which has no bearing on the solution of equation 190.

Leaving aside the nonlinear case for now, there is an obvious alternative to equation 192:

$$\partial_t \underline{v} = (2A+C)^T (\nabla_{\underline{v}} E) = -(2A+C)^T(2A+C)\underline{v} - (2A+C)^T\underline{x} \tag{193}$$

In this case, the dynamic matrix multiplying $\underline{v}$, $-(2A+C)^T(2A+C)$, is obviously a negative semidefinite symmetric matrix! Therefore, from linear systems analysis, we may be certain that this process is stable, for any A and C. Furthermore, unless 2A+C is singular, the condition for equilibrium is still equivalent to equation 190. In the case where 2A+C is singular, there <u>is</u> no unique equilibrium solution for the GMT problem, as we can see from equation 191; in that case, we are guaranteed that $\underline{v}$ will converge to <u>one</u> of the equilibria, all of which satisfy the first-order conditions for a minimum (equation 190).

How can we generalize equation 193 to the nonlinear/neural case? Recall from section 3.2 that backpropagation corresponds to premultiplying a vector by a transpose Jacobian; equation 193 has a suggestive similarity to equation 10. But in this case, we cannot just backpropagate through the nonlinear function or network which represents $E(\underline{v},\underline{w})$, because its output is only a scalar. However, we can derive equation 193 as the linear/quadratic special case of:

$$\partial_t \underline{v} = -\nabla_t (|\nabla_p E|^2) , \tag{194}$$

where $\nabla_p$ is the partial gradient and $\nabla_t$ the total gradient, as defined in section 4.3.2. The derivatives in equation 194 can be calculated efficiently, in a parallel distributed manner, by use of <u>second-order backpropagation</u> [8,ch.10]. Again, in the linear/quadratic case, those calculations reduce to equation 193.

As an alternative, we can avoid the need for second-order backpropagation by training an auxiliary network to estimate $\nabla_p E$. In other words, at each time t, we insert **v** and **w** into the network which calculates E. We can use first-order backpropagation to calculate $\nabla_p E$ efficiently, even if E is not a neural network [26]. We can then train an auxiliary network, using any form of supervised learning (like LMS), to input **v** and **w** and output a prediction or estimate of $\nabla_p E$. We can then use first-order backpropagation through that auxiliary network in order to calculate the derivatives used in equation 194.

It seems reasonably obvious that this variation should be stable in the linear-quadratic case, just like equation 193. If the auxiliary network (just a linear function, a matrix and a vector) is adapted by LMS, it converges to 2A+C" in a q-stable way, independent of the concurrent adaptation of **v**. The errors of this auxiliary model become as small as one likes, as time moves on. Thus the gap between this system and equation 193, in adapting **v**, also goes to zero. For any given estimate of 2A+C, the dynamics of **v** are stable, if one bases the adaptation of **v** entirely on the auxiliary network. (Alternatively, we could initialize the backpropagation through the auxiliary network based on the actual $\nabla_p E$ calculated by backpropagating through E.) Also note that this network is used to calculate derivatives which are summed over **v** and **w**; there is no need for special training methods which ensure that the network can disentangle the two.

In theory, one could also adapt an auxiliary network to approximate $\nabla_t E$. This would not eliminate the need for using second-order backpropagation in implementing or approximating equation 194.

The discussion above all concerns a gradient-based approach. For very difficult problems in the GMT family, it may be necessary to embed this kind of approach within a more global stochastic approach, to find the right basin of attraction. This would involve the notion of Brain-Like Stochastic Search mentioned in section 5.3. The details of that extension are beyond the scope of this paper. It should be noted, however, that there is a connection between local minimum problems and the particular version of "the chattering problem" discussed by Bertsekas [24]; one important way to reduce these problems is to use higher-order designs, as discussed in section 5.3.

**8.2.3. An Approach to Ensuring Convergence of Energy-Economy Models**

The approach in section 8.2.2 suggests a novel way to ensure convergence in large-scale models built out of nonlinear algebraic equations. In [16], I analyzed the problem of trying to ensure convergence of the official DOE model of long-term energy supply and demand across all sector. (This summarized a large DOE effort, in which I received support under contract from the Oak Ridge National Laboratory and the National Bureau for Economic Research at MIT.) For huge nonlinear models of this kind, achieving convergence is usually a matter of trial and error. There was no obvious way to really ensure convergence mathematically, even using the latest methods then available [16]; for example, methods based on conjugate gradients required an assumption of symmetric matrices, unlike the nonsymmetric Jacobians which these models possess.

Models of this sort are usually expressed in the form:

$$\underline{x} = \underline{f}(\underline{x}) , \qquad (195)$$

where $f_i$ is usually a formula "predicting" $x_i$ as a function of a few related variables. The naive way to converge such a model is by the usual update procedure:

$$\underline{x}^{(n+1)} = \underline{f}(\underline{x}^{(n)}) , \qquad (196)$$

where $\underline{x}^{(n)}$ is the n-th estimate of $\underline{x}$. Sometimes the process converges; sometimes it does not. When it does not, it is common to use a relaxation procedure, which, in the case of infinitely small learning rates, may be represented as:

$$\partial_t \underline{x} = (\underline{f}(\underline{x}) - \underline{x}) \tag{197}$$

In the neighborhood of the equilibrium, the error $\underline{\delta}$ in estimating $\underline{x}$ behaves like:

$$\partial_t \underline{\delta} = (F - I)\underline{\delta}, \tag{198}$$

where F is the Jacobian of $\underline{f}$. Even with infinitely small learning rates, however, there is no guarantee that equation 197 will converge, because there is no guarantee that F-I meets the usual Hurwicz criteria. Explosive modes of error growth remain possible. As an alternative, we may try:

$$\partial_t \underline{x} = (I-F^T)(\underline{f}(\underline{x})-\underline{x}) = \underline{f}(\underline{x}) - \underline{x} - F^T(\underline{f}(\underline{x}) - \underline{x}) \tag{199}$$

This may be implemented by taking the usual update, given in equation 197, and stabilizing it by subtracting the result of backpropagating it through the model $\underline{f}(\underline{x})$. Backpropagation through such a model is normally very fast, because of its sparse nonlinear structure. See [8,ch.10] and [26] for information on how to backpropagate through a nonlinear structure which is not a neural network. (Backpropagation has also been reinvented and implemented in some fast automatic differentiation programs, which could be useful in such applications.)

The errors in estimating $\underline{x}$ in equation 199 obey the following dynamics near equilibrium:

$$\partial_t \underline{\delta} = - (F^T-I)(F-I)\underline{\delta} \tag{200}$$

Like equation 193, this implies stable dynamics.

In high-performance implementations of this approach, we can use various neural network methods, such as the Adaptive Learning Rate (ALR) algorithm of [8,ch.3], to accelerate convergence. Or we could use the right-hand side of equation 199 in lieu of the gradient in a conjugate gradient method; see [22, ch.7] for a review of sophisticated conjugate gradient methods for highly nonlinear problems.

### 8.2.4. Extensions to Critic Adaptation and Stochastic (SGMT) Problems

Despite its potential value in GMT problems in general, equation 194 does not solve the problems described in sections 6 and 7.

The most obvious way to implement equation 194, in the linear stochastic case, would involve

$$\partial_t C = - \nabla_t (|\nabla_p E|^2) , \tag{201}$$

which, for an HDP type of method, entails:

$$\nabla_p E = -e x_i x_j \tag{202a}$$

$$(\nabla_p E)^2 = e^2 |\underline{x}|^4 \tag{202b}$$

$$<\partial_t C> = -<\nabla_t (|\nabla_p E|^2)> = - |\underline{x}|^4 <\nabla_t (e^2)>, \tag{202c}$$

which in turn reduces to HDPG, in effect. Section 8.2.1 assumed a <u>deterministic</u> GMT problem, in effect; here, the presence of noise invalidates a simple sampling-based
generalization of equation 194.

Another way to view this situation is to say, in the stochastic case, we really want to implement:

$$\partial_t C = -\nabla_t (<\nabla_p E>^2) ,  \qquad (203)$$

bearing in mind that $<\nabla_p E>^2$ is not the same as $<(\nabla_p E)^2>$. The Williams/Baird notion of "Bellman residual errors" [50,51] is related to this idea. The problem here is how to implement this idea, in the general case, where $<\nabla_p E>$ is not known except by some kind of sampling approach, which would interact with the adaptation process.

For the general Stochastic GMT (SGMT) problem, there is a straightforward solution to this difficulty. By using the <u>auxiliary network</u> approach of section 8.2.1, we automatically train an auxiliary network which outputs $<\nabla_p E>$ as a function of **v** and **w**. In this case, the approach based on an auxiliary network will converge to the right answer, even though the original approach based on second-order backpropagation does not. This provides what is perhaps the best general-purpose solution to the SGMT problem. As with GMT, we have a choice between using the auxiliary network itself, or backpropagation through E, in order to compute the derivatives fed back to the output of the auxiliary network, which initialize the process of backpropagation through the auxiliary network. (Also, as indicated in equation 194, **v** should be updated based on the <u>sum</u> of the derivatives fed back to **v** and the derivatives fed back to **w** in that final pass of backpropagation.)

Unfortunately, this is still not acceptable as a solution to the Critic adaptation problem. In the linear-quadratic case, the Critic matrix C would contain on the order of $n^2$ weights (actually, N' weights). The auxiliary network in this case would also input **v** and **w** -- again, on the order of $n^2$ inputs. Thus the auxiliary network would have on the order of $n^2$ inputs and $n^2$ outputs. It would be a matrix with $O(n^4)$ elements. This is not consistent with the stringent cost constraints which define parallel distributed processing or neural networks.
In addition, the need to estimate $n^4$ unknown parameters has negative effects on performance, not just cost., There is no straightforward and consistent way to solve the cost problem with this approach. In designing Critic adaptation rules, we must exploit more specific features
of the specific problem at hand.

## **8.3. An Alternative Analysis of the Problem**

This subsection will provide an alternative view of the nature of the Critic convergence problem. This alternative view will not really contradict the discussion above; it simply builds upon it.

One way to respond to section 7 is to enhance stability by trying to <u>recode</u> the estimated state vector, so as to make the dynamic matrix A closer to symmetric.
This would require some sort of recurrent coding process, to account for oscillatory modes in the original system. This is analogous to "reification" in physics, where a dynamic matrix can be represented in Hermitian form after a change in coordinate representation [78].
Reification approaches might be useful here; however, they would not eliminate the need to develop more robust Critic adaptation as well. At best, they could complement the effort to improve Critic adaptation rules as such. At "worst," they could become useless and obsolete, if new Critic adaptation rules do not work better in symmetric plants.

Another obvious approach is to try to detect and modify "bad directions," as proposed, for example, by Baird [51]. But there are great difficulties in detecting what is a "bad direction." The total gradient $\nabla_t$ should not be used as a standard for what is a good direction, since it does not point towards a minimum of what Baird calls "Bellman residual errors." The basic problem here lies in <u>finding</u> some vector which truly points in (or defines) a "good" direction.

Intuitively, in an HDP style of method, our real goal is to converge to a solution in which:
$$<e> = 0, \qquad (204)$$

in every state **x**, where the expectation is taken only over the random vector **w** previously discussed. (Some researchers have talked about the great difficulty of calculating derivatives through stochastic systems, and the breakthroughs of the early 1990's in that area; however, backpropagation through a function **f**(**x**,**w**), treating **w** as a fixed input, is extremely straightforward. It was described in excruciatingly explicit detail years ago [65].)
Thus in an ideal world, we would indeed want to implement:

$$\partial_t W = -\nabla_t (<e>^2) = -2<e><\nabla_t e> \qquad (205)$$

Our problem is that this is not the same as $<e\nabla_t e>$, because of statistical correlation effects between e and $\nabla_t e$. Thus if we adapt W in proportion to $e\nabla_t e$, as in GDHP, we are sampling $<e\nabla_t e>$, which is not what we want. If we adapt

$$\partial_t W = -\nabla_p (<e>^2) = -2<e><\nabla_p e> = -2<e\nabla_p e>, \qquad (206)$$

as in pure HDP, we <u>can</u> do ordinary sampling, because $\nabla_p e$ does not depend on the random vector **w**; the statistical correlation effects disappear, so that we can get what we want by adapting in real time in proportion to $e\nabla_p e$. Equation 206, like equation 205, reaches equilibrium at the correct desired state, the state where $<e> = 0$, <u>when it reaches equilibrium</u>.
The problem with HDP is that $\nabla_p e$ does not always take us in a valid direction to <u>reach</u> its equilibrium. In summary, in pure HDP, we are adapting the weights W in proportion to $e(\nabla_p e)$, where "e" gives us the real error signal and the guarantee of correct equilibrium, while the $\nabla_p e$ term provides the convergence or stability properties, such as they are.

## 9. New Methods for Critic Adaptation

This section will discuss twelve alternative new variants of HDP, of which one (the two-sample method [50,51]) has been previously published. The corresponding variants of DHP and GDHP will also be discussed.

The twelve variants actually consist of four core variants -- all of which solve the main problem of sections 6 and 7 -- and two ways of modifying them.

This section will begin by discussing the core variants, in section 9.1. Section 9.2 will discuss the "sailboat" and "weighted sailboat" variants of these. Section 9.3 will discuss how to implement similar variants of DHP and GDHP.

### 9.1. Four New Core Variations of HDP

The equilibrium problems with HDPG discussed in section 6 were previously discussed, in a more primitive example, back in 1990 [17]. Baird [51] suggested in 1995 that these problems could be solved, in effect, by trying to implement equation 205. To solve the statistical correlation problem, which results in $\langle e \nabla_t e \rangle \neq \langle e \rangle \langle \nabla_t e \rangle$, they proposed that "e" be estimated in the usual manner, but that "$\nabla_t e$" be estimated by simulating a <u>different</u> state vector **R**(t+1), based on a different random vector, $\underline{w}_2$, using the available plant model **R**(t+1)=**f**(**R**(t),**w**(t)). (Recall from section 4.3.2 that all references to a "plant model" in this section actually refer to an augmented plant model; however, when I propose ways to adapt such an augmented model, in this section, I am referring to adaptation only of weights in the Model network proper.) In his scheme,

$$\langle e_1 \nabla_t e_2 \rangle = \langle e_1 \rangle \langle \nabla_t e_2 \rangle = \langle e \rangle \langle \nabla_t e \rangle, \tag{207}$$

because there are no correlations between the two different random vectors $\underline{w}_1$ and $\underline{w}_2$.

So far as I know, the only actual implementation of this idea -- which was proposed only briefly in a kind of side section of [51] -- was by Prokhorov [15]. Prokhorov has reported that the high level of random noise in this algorithm, due to the use of two unrelated random vectors, leads to very slow convergence, particularly near the equilibrium state. Nevertheless, this is an example of a method which solves the problem of sections 6 and 7 in a formal sense. It would converge to a correct equilibrium, the equilibrium where $\langle e \rangle = 0$. It would possess q-stability, in the stochastic sense discussed in section 7.1, because the expected dynamics of M follow the dynamics of M for HDPG in the deterministic case.

This first correct variation of HDP may be called "HDP2," to highlight the use of <u>two</u> random vectors in parallel.

An alternative approach is to train a <u>deterministic forecasting model</u> $\underline{f}'$ (**R**(t)) to predict the <u>expectation value</u> of **R**(t+1). (Actually, this is the most common way to train neural net system identifiers today; few researchers seem to be aware of the stochastic alternatives [8,ch.13;75].) One can adapt W in proportion to $e \nabla_t e'$, where:

(1) the "e" on the left is calculated in the usual way, based on the sampled value of $\hat{J}$(t+1);

(2) $\nabla_t e'$ is calculated by backpropagation through:

$$e' = (\hat{J}(\mathbf{R}(t,W)) - (\hat{J}(\underline{f}'(\mathbf{R}(t)),W)+U)) \tag{208}$$

This results in:

$$\langle \partial_t W \rangle = -\langle e \rangle \nabla_t e', \tag{209}$$

which has <u>exactly the correct equilibrium</u>, $\langle e \rangle = 0$, exactly like equation 206. In other words, this method should have exactly the same equilibrium as HDP, with the same degree of noise filtering as in HDP. The replacement of $\nabla_p e$ by $\nabla_t e'$ simply leads to a direction of movement which is much more likely to avoid convergence problems. This variation on HDP may be called HDP0, in order to emphasize the lack (0) of statistical sampling in determining derivatives of e.

HDP0 and HDP2 both require a heavy use of a system identification component, such as the adaptive estimates of A in the linear/quadratic case. In the terminology of section 4.3, HDP0 replaces equations 25 and 26 by:

$$\Delta W = -2\alpha e(\nabla_t e') = 2\alpha e(\nabla \hat{J}(\mathbf{R}(t)) - (1/(1+r))\nabla \hat{J}(\underline{f}'(\mathbf{R}(t)))) \tag{210}$$

In the linear/stochastic case, we still compute e by sampling (as implied in equation 75), but we adapt the weights according to equation 71, using the <u>estimated value</u> of the matrix A to perform the calculation in equation 71. **This is my first recommended procedure for adapting Critics** **in the linear/quadratic case, in order to permit the development of a universal adaptive controller**. (For the second, see the end of section 9.3).

There are certain obvious tradeoffs in the use of HDP0. Like IAC itself (section 3), it depends heavily on the estimate of A. From the viewpoint of long-term stability theorems, in the spirit of Narendra [1], this is not an essential problem, because the estimates of A should eventually become as accurate as we like, if we pay the usual amount of attention to issues like persistence of excitation and like focusing on estimation errors only in directions which are actually relevant to policy. In practical situations, the adaptation of a suitable problem-dependent Critic should allow faster, more effective adaptation by the other parts of the overall control system, as in some of the aerospace simulations published by Balakrishnan. One could go even further by dispensing with the sampling altogether' one could use estimates of A and Q to estimate <e> directly from the expected value of equation 75; however, that variation -- unlike HDP0 -- does not work in a low-cost, consistent way in the nonlinear stochastic case.

Again, HDP0 is just as exact as HDP proper, but more reliable in terms of convergence. In a formal statistical sense, HDP2 may be even more reliable in its convergence, because it implies an <u>exact</u> symmetry (on average). But the discussion of section 8.1 suggests that <u>exact</u> symmetry is probably not necessary even for unconditional stability.

The greater filtering of noise in HDP0 compared with HDP2 should be a tremendous advantage in practical terms. One might imagine trying to filter the "e" signal as well, to some degree; however, the Adaptive Learning Rate (ALR) [8,ch3] family of methods already achieves the same effect.

Nevertheless, there are two further variants of HDP0 which may be worth exploring, in cases where precise symmetry in convergence is worth the cost.

First, in HDP02 -- a second-order variation of HDP0 -- we train the network $\underline{\mathbf{f}}'$ as follows (after an initial period of adapting it to predict $\underline{\mathbf{R}}(t+1)$). At each time t, we obtain a sample of $\underline{\mathbf{R}}(t+1)$ in one of the usual ways -- either by simulating a random vector $\underline{\mathbf{w}}$ inserted into a stochastic model $\underline{\mathbf{f}}(\underline{\mathbf{R}}(t),\underline{\mathbf{w}})$ or by actually observing $\underline{\mathbf{R}}(t+1)$ and estimating $\underline{\mathbf{w}}$. We then backpropagate the usual error e through $\hat{J}(\underline{\mathbf{R}}(t+1))$ in the usual way, back to the weights in the Critic network. Let us call the resulting gradient of e with respect to the Critic weights "$\mathbf{g}_1$". Then we backpropagate e' through $\hat{J}(\underline{\mathbf{f}}'(\underline{\mathbf{R}}(t))$, back to the same weights in the same Critic network. Let us call the second set of derivatives $\mathbf{g}_2$. Notice that $\mathbf{g}_1$ and $\mathbf{g}_2$ are both gradients with respect to weights <u>in the Critic network</u>; each component of these gradients may be physically located "inside" the corresponding weight or synapse of the Critic, in a parallel distributed implementation. Then we adapt the weights of $\underline{\mathbf{f}}'$ so as to minimize:

$$E_g = |\mathbf{g}_1 - \mathbf{g}_2|^2 \tag{211}$$

In other words, we propagate these "g" errors back through the calculations which led to $\mathbf{g}_2$, using second-order backpropagation, in order to calculate the derivatives of $E_g$ with respect to the weights in $\underline{\mathbf{f}}'$. This procedure trains $\underline{\mathbf{f}}'$ to give us the most exact possible filtered, unbiased estimates of $<\nabla_t e>$. In a practical implementation, we would actually adapt the weights of $\underline{\mathbf{f}}'$ based on a weighted sum of these second-order derivatives and the simple derivatives of error in predicting $\underline{\mathbf{R}}(t+1)$, with the relative weights of the two types of error adjusted gradually over time.

Finally, there is a variation of these methods which becomes accessible only after the implementation of more complex designs like SEDP [8,ch.13] and the 3-brain architecture [46]. In such systems, vectors become available which represent probability distributions rather than states as such. In such systems, we may use these vectors as inputs or outputs of $\underline{\mathbf{f}}'$ instead of using an estimated state vector; we may also use them in developing partially closed form estimates of <e>, in order to reduce random sampling error. As in SEDP, we may use a neurally-based weighted or stratified sampling strategy instead of a random sampling strategy for the remaining sampling requirements. Again, these details are well beyond the present state of the art; however, they will become more and more relevant as we move towards the construction of true intelligent systems as proposed in section 5.3.

## 9.2. "Sailboat" Variations of These Methods

All four of the methods discussed in section 9.1 calculate directions of movement, **δ**, for the weights, **W**. These directions of movement are truly "good" directions, insofar as they point (on average) to the correct equilibrium Critic weights. These directions are all based on a simultaneous consideration of $\hat{J}(t+1)$ and $\hat{J}(t)$, as is appropriate when controlling an ergodic sort of process where we bounce back and forth forever between the same set of states.
 On the other hand, traditional dynamic programming works best when we can work backwards from a terminal state, through to intermediate states, through to starting states. Bertsekas calls this kind of situation a "layered" state space [24]. HDP implements this kind of backwards flow of information, from time t+1 to time t. In "layered" situations, we might imagine that HDP would perform better than total-gradient methods, especially if the networks are constructed such that the changes in weights W do not affect $\hat{J}(t+1)$ very much anyway.
 Unfortunately, HDP (and DHP or GDHP) is not a perfect way to capture the potential benefits of a layered state space. If the Critic involves any degree of global generalization, it is difficult to ensure that the weight changes do not affect $\hat{J}(t+1)$ significantly. Also, the layering may not be perfect or exact. As an example, the autolander problem studied by Dolmatova and myself [53] appears to involve a monotonic sort of progression from start to finish, but we found that convergence and stability issues like those discussed in sections 7 and 8 were still significant.
 There is an alternative way to try to exploit the special properties of layered state spaces. One can simply project the "correct direction" -- as calculated by one of the methods discussed in section 9.1 -- onto the line defined by $\nabla_p E$. (For HDP, this is simply the line defined by $\nabla_p e$.) This is the "sailboat method." It is that simple. I call it the sailboat method because $\nabla_p E$ acts like the rudder of a boat, defining the line of motion, while the correct gradient acts like the wind, defining the degree of motion along that line. But in this case, the motion may be forwards or backwards. In order to maintain the guarantees of consistency and stability, we must allow the change in weights to go opposite to $-\nabla_p E$,
when the correct gradient tells us to do so.
 A further variation of this is to use a weighted sum of the correct gradient and the sailboat gradient. The weight on each would be simply an adapted learning rate on each, calculated by the Adaptive Learning Rate (ALR) algorithm of [8,ch.3] or something similar.
In effect, this variation would adaptively decide what the right combination of methods is for the particular environment or plant under consideration.
 The higher-order multiple-model Critic designs discussed in [9,13,46] would naturally tend to create a layered condition within each "decision block" or region.
Therefore, the sailboat methods -- pure and weighted -- could be of great value when combined with that approach. This suggests that true intelligent systems will use a relatively smoother sailing adaptation for lower-level decisions (encased in a layered region) but may have greater difficulties with stability and convergence when dealing with the very highest levels of decisions and emotional priorities. This is consistent with what we see in humans.

## 9.3. Extensions to DHP and GDHP

This section will discuss how to extend the four methods of sections 9.1 to DHP and GDHP. The sailboat versions are all straightforward, based on the discussion of the previous section.

For the two-sample method, we define DHP2 by replacing equation 28 by:

$$\lambda_i^*(t) = \frac{\partial U(t)}{\partial R_i} + \frac{1}{1+r} \sum_j \hat{\lambda}_j(t+1) \cdot \frac{\partial f_j(\underline{R}(t), \underline{w_2}(t))}{\partial R_i(t)} \quad (212)$$

where $\hat{\lambda}_j(t+1)$ is calculated in the usual way (based on a random vector $\underline{w}_1(t)$, explicitly simulated or imputed from observation) but the derivative on the right is calculated for $\underline{f}$ inputting a different random vector, $\underline{w}_2$. For GDHP2, we now adapt the weights W in proportion to the following estimate of the gradient of equation 32:

$$\Omega_0 e_0 (\nabla_t e_0') + \underline{e}^T \Omega \nabla_t \underline{e}' \quad , \quad (213)$$

where $e_0$ and $\underline{e}$ are computed in the usual way, and where $e_0'$ and $\underline{e}'$ are computed like $e_0$ and $\underline{e}$ except that they are based on the simulated random vector $\underline{w}_2$ rather than actual observation of $\underline{R}(t+1)$ or $\underline{w}_1$. As in ordinary GDHP, second-order backpropagation is used to calculate this gradient.

For the direct method, we define DHP0 by:

$$\lambda_i^*(t) = \frac{\partial U(t)}{\partial R_i} + \frac{1}{1+r} \sum_j \hat{\lambda}_j(t+1) \cdot \frac{\partial f_j'(\underline{R}(t))}{\partial R_i(t)} \quad (214)$$

where $\hat{\lambda}_j(t+1)$ is again calculated in the usual way, and $\underline{f}'$ is the same auxiliary network discussed in section 9.1 for HDP0. DHP02 also uses equation 214, but adapts $\underline{f}'$ using second-order backpropagation, based on the gradient of E (as defined in equation 27) with respect to the Critic weights; the details follow the discussion of HDP02 exactly, although the calculations are different simply because the calculations for second-order backpropagation are slightly different when applied to this different error function E. GDHP0 and GDHP02 is again based on equation 213, but with a different definition of $e_0'$ and $\underline{e}'$; in this case, these errors are calculated based on $\underline{R}(t+1) = \underline{f}'(\underline{R}(t))$, for the same function $\underline{f}'$ we have discussed for HDP0 and HDP02.

For DHP and GDHP, there are further variants which involve using vectors which represent probability distributions rather than particular states and so on, exactly as discussed for HDP in section 9.1.

**For the development of universal adaptive controllers in the linear-quadratic case, I recommend consideration of GDHP0 as the most promising alternative to HDP0 at present.** For simplicity, we may use $\Omega_0=0$ and $\Omega=I$ for now. GDHP0 with $\Omega_0=0$, unlike HDP0, avoids the need to think about the $U_0$ term in the quadratic case. By exploiting symmetry and exploiting additional information, GDHP0 may well permit more rapid adaptation than the alternative methods which have nonlinear, stochastic generalizations.